**Efficient Photocatalytic CO$_2$ Reduction to C$_{2+}$ Products with Pt$_{1-x}$Pd$_x$Sn$_4$ Dirac Nodal Arc Semimetal**

*Kangwang Wang, Jie Zhan, Jun Liu, Zaichen Xiang, Wanyi Zhang, Lingyong Zeng, Kai Yan\*, Yan Sun\*, and Huixia Luo\**

K. Wang, J. Liu, Z. Xiang, W. Zhang, L. Zeng, Prof. H. Luo
School of Materials Science and Engineering, State Key Laboratory of Optoelectronic Materials and Technologies, Guangdong Provincial Key Laboratory of Magnetoelectric Physics and Devices, Key Lab of Polymer Composite & Functional Materials,
Sun Yat-sen University,
Guangzhou 510275, China
E-mail: luohx7@mail.sysu.edu.cn

Prof. K. Yan
School of Environmental Science and Engineering,
Sun Yat-sen University,
Guangzhou 510275, China
E-mail: yank9@mail.sysu.edu.cn

J. Zhan, Prof. Y. Sun
Shenyang National Laboratory for Materials Science, Institute of Metal Research,
Chinese Academy of Sciences,
Shenyang 110016, China
School of Materials Science and Engineering,
University of Science and Technology of China, Shenyang 110016, China
E-mail: sunyan@imr.ac.cn

**Abstract**

The photochemical CO$_2$ reduction reaction (CRR) represents a zero-carbon pathway for converting CO$_2$ into value-added chemicals, yet its industrial implementation has been constrained by low selectivity and product diversity. Dirac nodal arc semimetals characterized by ultrahigh carrier mobility (>25,000 cm$^2\cdot$V$^{-1}\cdot$s$^{-1}$) offer a promising platform to search for efficient catalysts for CO$_2$ conversion. Herein, we demonstrate that strategic Pt incorporation into PdSn$_4$ optimizes the electronic structure and carrier dynamics of this Dirac semimetal. Experimental and theoretical analyses reveal that the resulting Pd−Sn−Pt local electronic structure redistributes charge density around Pd and Pt atoms, which facilitates C−C coupling via *OC−COH and *OC−CHOH intermediates and enhances carrier mobility by 40% versus the pristine PdSn$_4$ single crystal. The optimized Pd$_{0.4}$Pt$_{0.6}$Sn$_4$ single crystal achieves C$_2$H$_4$ (i) formation rate of 328 μmol·g$^{-1}$·h$^{-1}$; (ii) product selectivity of 73.1%; (iii) electron-based selectivity of 89%. This work establishes electronic-structure-tunable Dirac semimetals as a new paradigm for multi-carbon photochemical CO$_2$ reduction, providing a design strategy for next-generation photocatalysts.

**1. Introduction**

Nowadays, energy consumption is still dominated by nonrenewable fuels such as fossil fuels, and the challenge of insufficient energy supply is becoming increasingly prominent. Thus, it is crucial to explore innovative and substitute energy sources.[1] Among various strategies for obtaining renewable hydrocarbon resources, photocatalytic technology is likely to be the most cost-effective and operationally simple approach. Developing highly efficient photocatalysts for converting CO$_2$ into chemical fuels is significant for energy sustainability and carbon resource utilization.[2] From this perspective, efficient photocatalytic CO$_2$ reduction reaction (CRR) mainly depends on highly active photocatalysts and feasible photocatalytic reaction pathways and mechanisms.[3] Compared with the mono-carbon (C$_1$) products (such as carbon monoxide (CO), methanol (CH$_3$OH), formic acid (HCOOH), methane (CH$_4$), etc.), the multi-carbon (C$_{2+}$) products (such as acetylene (C$_2$H$_2$), ethylene (C$_2$H$_4$), ethanol (CH$_3$CH$_2$OH), propanol (CH$_3$CH$_2$CH$_2$OH), etc.) have higher energy density and market value.[4] Theoretically, under mild conditions, in the process of CO$_2$ photoreduction to C$_{2+}$ products, the key limiting step is the C−C coupling process among those hydrogenation intermediates (for instance, CO*,

*COOH, *CHO, *COH, *OCCO, and *OCCOH, etc.), as the kinetic energy barrier of this process is very high.[2] Nonetheless, due to the complex multi-electron reduction reaction pathways, achieving satisfactory $C_{2+}$ selectivity and internal quantum efficiency (*IQE*) remains a severe challenge.

Recent observations indicate that topological electronic states can mediate charge transfer from substrates to adsorbates, thereby altering their interfacial chemical bonds between the substrate and the adsorbate. Based on this mechanism, topological catalysis has emerged as an increasingly recognized and promising direction in catalysis and energy conversion research. Moreover, in most topological semimetal materials, the linear bands display a phenomenon of crossing, which results in extremely high electrical conductivity and charge carrier mobility at room temperature.[5] For instance, $PtSn_4$ semimetal is an extremely attractive topological material platform, featuring a nontrivial topological surface state (TSSs) and Dirac node arcs.[6] As a hydrogen evolution reaction (HER) catalyst, the turnover frequencies (TOFs) of $PtSn_4$ are 1.5 $H_2 \cdot s^{-1}$ at 100 mV.[7] Noticeably, another Dirac nodal arc semimetal $PdSn_4$ with the same crystal structure as $PtSn_4$, was found to have a Tafel slope of 83 $mV \cdot dec^{-1}$ in the HER with an overpotential of 50 mV at a current density of 10 $mA \cdot cm^{-2}$.[8] In addition, $PdSn_4$ semimetal, with a crystal structure similar to $PtSn_4$, shows superior robustness to CO and higher stability in aqueous media.[8] Kong et al. recently demonstrated that nontrivial TSSs can be precisely tuned through nanosheet thickness and magnetic field control, offering direct experimental proof of their contribution to the CRR.[9] Despite this advance, a detailed understanding of the operative mechanism of the nontrivial TSSs and the specific electronic structural dynamics involved is still lacking.

As mentioned above, we employ the topological Dirac nodal arc semimetal $Pd_{1-x}Pt_xSn_4$ (*x* = 0.0, 0.4, 0.6, and 1.0) as the catalyst to investigate the relationship between the surface chemical reaction activity and the nontrivial TSSs in the CRR. Experimentally, we found that the composition exhibiting the highest carrier mobility delivers the highest CRR performance.

Consequently, the optimal composition $Pd_{0.4}Pt_{0.6}Sn_4$ single crystal achieves an exceptional $C_2H_4$ production rate of 328 μmol·g$^{-1}$·h$^{-1}$, with $C_2H_4$ selectivity of 73.1% and $C_2H_4$ electron-based selectivity of 89%. This performance surpasses state-of-the-art Sn-based nanostructured photocatalysts despite their considerable surface areas. Furthermore, a series of in-situ characterization techniques combined with theoretical calculations reveals that Pd−Sn−Pt multi-active sites, arising from the local atomic structure, lower the reaction barrier for forming the crucial intermediates *OC−COH and *OC−CHOH. Additionally, advanced photoelectric characterization and theoretical computations demonstrate that the CRR activity of $Pd_{1-x}Pt_xSn_4$ single crystals scales proportionally with carrier concentration. Our results establish an intrinsic link between Dirac nodal arc semimetals and photocatalysis, and introduce carrier concentration as a universal extrinsic parameter for exploring topological effects in catalysis.

## 2. Results and Discussion

### 2.1 Kinetics and Thermodynamics of Carriers

The origin of the excellent photocatalytic performance on $Pd_{0.4}Pt_{0.6}Sn_4$ single crystal was investigated by advanced photoelectric properties characterizations, including ultraviolet-visible diffuse reflectance spectra (UV-vis DRS), transient photocurrent, electrochemical impedance spectra (EIS), steady-state photoluminescence (PL), time-resolved photoluminescence (TRPL), and Hall resistivity measurements. The UV-vis DRS spectra indicate that the $Pd_{1-x}Pt_xSn_4$ single crystals show strong absorption from the ultraviolet to near-infrared field (Figure 1a). Meanwhile, the $Pd_{1-x}Pt_xSn_4$ single crystals broaden the light absorption range, promoting the absorption and utilization of solar energy.[10] The EIS analysis shows that the charge-transfer resistance ($R_{ct}$) of the $PdSn_4$, $PtSn_4$, $Pd_{0.4}Pt_{0.6}Sn_4$, and $Pd_{0.6}Pt_{0.4}Sn_4$ single crystals are approximately 95.98, 74.55, 40.65, and 68.04 Ω, respectively (Figure 1b), indicating that the $Pd_{0.6}Pt_{0.4}Sn_4$ single crystal has more favorable electron transfer and charge separation. We also conducted steady-state PL measurements to compare the photo-induced charge transfer and charge recombination loss in this $Pd_{1-x}Pt_xSn_4$. As shown in Figure 1c, the PL intensity of the $Pd_{0.4}Pt_{0.6}Sn_4$ single crystal is much lower than that of the $PdSn_4$, $PtSn_4$, and $Pd_{0.6}Pt_{0.4}Sn_4$ single crystals, respectively, indicating that the serious charge recombination in the $Pd_{0.4}Pt_{0.6}Sn_4$ single crystal is significantly suppressed.[11] The TRPL test was carried out to monitor and analyze the charge carrier dynamics on the nanosecond time scale. As shown in Figure 1d, the average carrier lifetime is 3.0200 ns for the $Pd_{0.4}Pt_{0.6}Sn_4$ single crystal, which is smaller than that of the $PdSn_4$ (20.3602 ns) and $PtSn_4$ (18.3581 ns) single crystals, respectively. Table S1 (Supporting Information) shows that as the proportion of Pt atoms increases, the average carrier lifetime gradually decreases (3.3617 ns for $Pd_{0.6}Pt_{0.4}Sn_4$). Meanwhile, the Hall effect measurement at 300 K indicates that the semi-metallic nature of $Pd_{0.4}Pt_{0.6}Sn_4$ single crystal with a carrier concentration of $1.01\times10^{29}$ $m^{-3}$ (Figure 1e) and a carrier mobility of $8.36\times10^{-3}$ $m^2\cdot V^{-1}\cdot s^{-1}$ (Figure 1f).[12] In addition, the $PdSn_4$ single crystal has a carrier concentration of $1.75\times10^{28}$ $m^{-3}$ and a carrier mobility of $9.41\times10^{-4}$ $m^2\cdot V^{-1}\cdot s^{-1}$, respectively.[13] The calculated value of the carrier diffusion lengths ($L_D$) ranges from 0.119 to 0.292 μm for $Pd_{0.4}Pt_{0.6}Sn_4$ single crystal, 0.049 to 0.231 μm for $PdSn_4$ single crystal, and 0.046 to 0.234 μm for $PtSn_4$ single crystal, respectively (Table S2, Supporting Information).[14] Therefore, the photoelectric properties characterizations indicate that the $Pd_{0.4}Pt_{0.6}Sn_4$ single crystal is more efficient for charge transfer, which clarifies why better photocatalytic CRR activities were obtained on the $Pd_{0.4}Pt_{0.6}Sn_4$ single crystal.

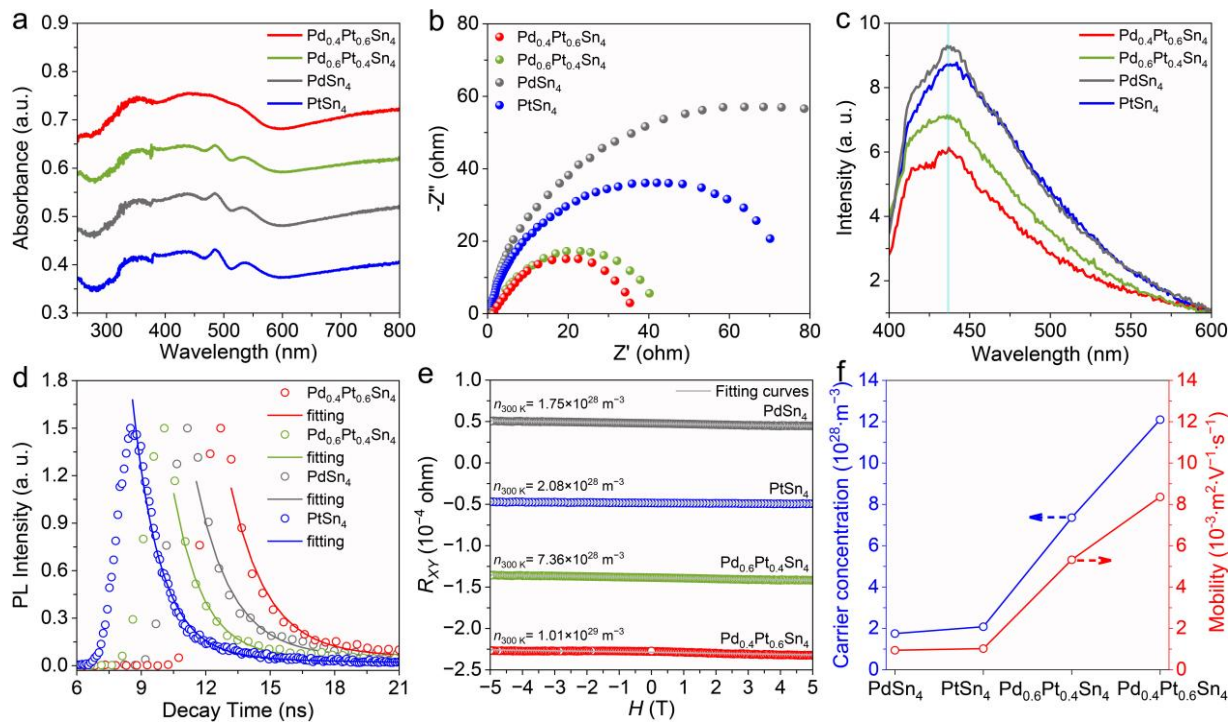

Figure 1. The kinetics and thermodynamics of carriers in designed photocatalysts. (a) UV-vis DRS, (b) EIS, (c) PL, and (d) TRPL spectra of the $PdSn_4$, $PtSn_4$, $Pd_{0.4}Pt_{0.6}Sn_4$, and $Pd_{0.6}Pt_{0.4}Sn_4$ single crystals. (e) Hall resistivity versus magnetic field of the $PdSn_4$, $PtSn_4$, $Pd_{0.6}Pt_{0.4}Sn_4$, and $Pd_{0.4}Pt_{0.6}Sn_4$ single crystals at 300 K. (f) Carrier concentration and mobility of the $PdSn_4$, $PtSn_4$, $Pd_{0.6}Pt_{0.4}Sn_4$, and $Pd_{0.4}Pt_{0.6}Sn_4$ single crystals.

## 2.2 Morphology and Structure Analysis

The X-ray diffraction (XRD) patterns taken on a mechanically cleaved crystal flake of a single crystal are displayed in Figure S1a. High intensity and sharp peaks are observed for only (002n) planes, i.e., (002), (004), (006), and (008). This shows that the synthesized crystals are grown unidirectionally along the $c$-axis, confirming the crystallinity of the identical. The phase purity of $PdSn_4$, $PtSn_4$, $Pd_{0.4}Pt_{0.6}Sn_4$, and $Pd_{0.4}Pt_{0.6}Sn_4$ single crystals was supported through the Rietveld refined powder XRD (PXRD) patterns (Figure S1b−c, Supporting Information). The obtained lattice parameters are $a$ = 6.3843 Å, $b$ = 6.4203 Å, $c$ = 11.3681 Å, and $\alpha = \beta = \gamma = 90°$. In the present article, the PXRD pattern of $Pd_{1-x}Pt_xSn_4$ further confirmed that $Pd_{0.4}Pt_{0.6}Sn_4$ exists in an orthorhombic phase and has the Aea2 space group symmetry. Notably, the relative intensity of the peaks in the single-crystal samples is different from that of the calculated powder pattern, which is mainly due to the preferential orientation of the ground layered structure powder. Atomic resolution high-angle annular dark-field scanning transmission electron microscopy (HAADF-STEM) images of $Pd_{0.4}Pt_{0.6}Sn_4$ single crystal (space group I41/Aea2) (Figure 2a,b) and $PdSn_4$ single crystal (space group I41/Aea2) (Figure S2, Supporting Information) show a zigzag crystalline structure along the [010] intermetallic facet. This observation is further corroborated by the intensity distribution in the regions of Line 1 and Line 2 (Figure 2b). This distribution shows that there are significant differences in the Z-line intensities between different atoms, and this difference is directly related to the atomic numbers of Pd and Pt elements (Figure 2c,d). The atomic arrangements are further reconfirmed by the corresponding crystal structure (Figure 2e). Simultaneously, two clear lattice distances of 3.33 and 3.15 Å are observed, matching strongly with its simulated crystalline phase structure (Figure S3, Supporting Information). In sharp comparison, the $PdSn_4$ single crystal holds a similar crystalline structure and meso-structure but a different atomic ratio. The selected area electron diffraction (SAED) patterns of the $Pd_{0.4}Pt_{0.6}Sn_4$ single crystal are displayed in Figure 2f, which can be indexed along the [010] directions, respectively. The atomic-resolved element mapping was further performed to investigate the distribution of Pt, Pd, and Sn atoms. Superposition of the Pt, Pd, and Sn element mappings combined with the corresponding HAADF image indicates the uniform distribution of Pt, Pd, and Sn atoms on the $Pd_{0.4}Pt_{0.6}Sn_4$ single crystal surface (Figure 2g). Furthuly, the atomic strain distribution patterns obtained through geometric phase analysis (GPA) in Figure 2h,i clearly indicate that the lattice distortion throughout the entire area is extremely severe, which is mainly attributed to the changes in the atomic radii of Pd and Pt.[15] The $Pd_{0.4}Pt_{0.6}Sn_4$ single crystal has the characteristics of local lattice distortion, which can efficiently tune the band structure. The detailed analysis results mentioned above indicate that the $Pd_{0.4}Pt_{0.6}Sn_4$ single crystal possesses high purity and high crystallinity, which can significantly highlight the influence of the single crystal electronic structure on the CRR catalytic activity.

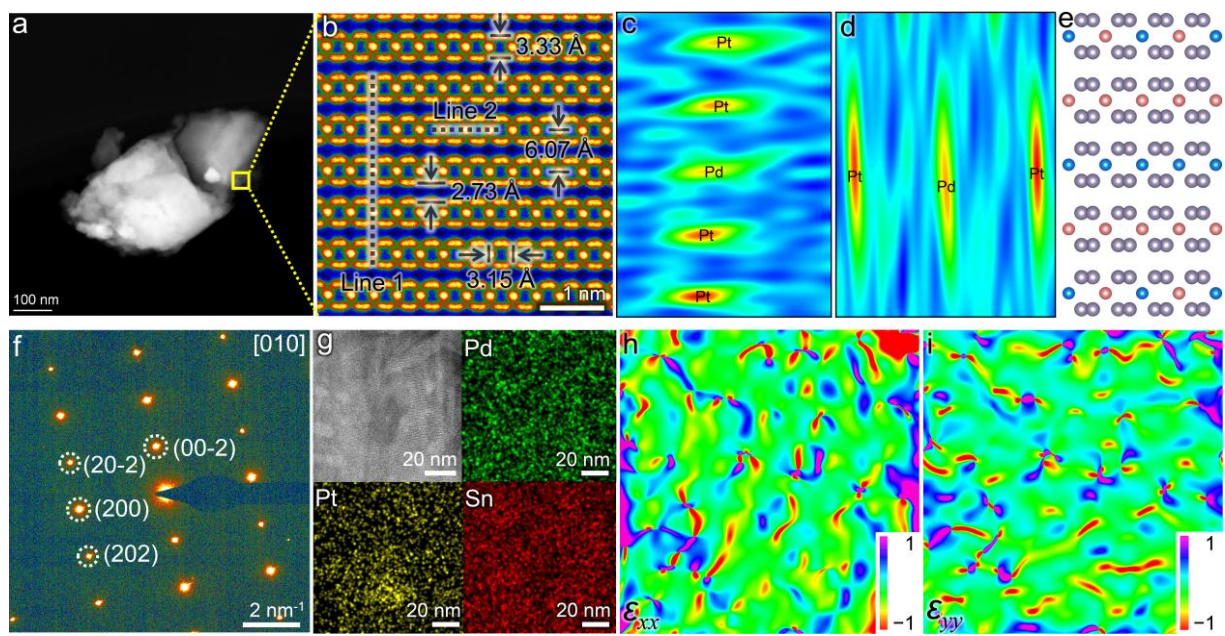

Figure 2. Crystal structure characterization of $Pd_{0.4}Pt_{0.6}Sn_4$ single crystal. (a,b) HAADF-STEM images of $Pd_{0.4}Pt_{0.6}Sn_4$ single crystal. The intensity profile along (c) Line 1 and (d) Line 2 is indicated in panel (b). (e) The optimized crystal structure of $Pd_{0.4}Pt_{0.6}Sn_4$ single crystal from the [010] direction (the gray, blue, and red spheres represent Sn, Pd, and Pt atoms). (f) SEAD pattern taken from the [010] direction, and (g) EDS mapping of $Pd_{0.4}Pt_{0.6}Sn_4$ single crystal. (h, i) GPA analysis based on the HAADF image (d).

## 2.3 Electronic Structure Analysis

Since the valence state of the active site plays a crucial role in the catalytic activity of the catalysts, X-ray absorption spectroscopy (XAS) was carried out to investigate the chemical states and coordination environments of $PdSn_4$, $PtSn_4$, and $Pd_{0.4}Pt_{0.6}Sn_4$ single crystals at the atomic level. Figure S4a−c (Supporting Information) depicts the X-ray Absorption Near-edge Fine Structure (XANES) curves of the Pd $K$-edge, Pt $L_3$-edge, and Sn $K$-edge, respectively, which reveal the pre-edge features based on the absorption edge or the location of white lines. The absorption edges of the Pd $K$-edge for the $Pd_{0.4}Pt_{0.6}Sn_4$ and $PdSn_4$ single crystals are located between those of the Pd foil ref. and PdO ref. (Figure S4a, Supporting Information), meaning the existence of positively charged Pd species. In addition, the absorption edge of $Pd_{0.4}Pt_{0.6}Sn_4$ single crystals is closer to that of Pd foil ref., suggesting a lower chemical state with deeper reduction.[16] Note that the electronic structure of Pd atoms is finely modulated after the doping with Pt atoms. The shapes of the Pt $L_3$-edge for the $Pd_{0.4}Pt_{0.6}Sn_4$ and $PtSn_4$ single crystals show that the positions of the absorption threshold of $Pd_{0.4}Pt_{0.6}Sn_4$ and $PtSn_4$ single crystals are situated in the Pt foil ref. and $PtO_2$ ref. (Figure S4b, Supporting Information), whereas the white line peak intensity at ~11567 eV is much weaker than the $PtO_2$ ref., revealing that the Pt atoms in the material are mainly in the metallic state.[17] A positive shift of the absorption edge position is found for the Sn $K$-edge after interacting with Pd and Pt atoms, compared with that of the Sn foil ref. (Figure S4c, Supporting Information), illustrating the electron density transfer from the Sn atom to the Pd and Pt atoms. In the Pt $L_3$-edge XANES of the $Pd_{0.4}Pt_{0.6}Sn_4$ single crystal, the intensity of the white line peak is slightly lower than that of the Pt foil ref. and $PtSn_4$ single crystal, indicating that the valence state of the Pt element in $Pd_{0.4}Pt_{0.6}Sn_4$ single crystal is in a reduced state.[18] Astonishingly, a slight increasing of the white line peak intensity of Sn $K$-edge for the $Pd_{0.4}Pt_{0.6}Sn_4$ single crystal compared to the $PdSn_4$ and $PtSn_4$ single crystals also indicates that the electron density of Pd and Pt atoms in the $Pd_{0.4}Pt_{0.6}Sn_4$ single crystal is relatively high, further suggesting that Sn transfers electrons to Pd and Pt.[19]

To unveil the local coordination structures of the Pd, Pt, and Sn atoms in the $Pd_{0.4}Pt_{0.6}Sn_4$ single crystal, we further performed the Fourier-transformed extended X-ray absorption fine structure (FT-EXAFS) spectra, where a prominent peak located at 2.49 Å is assigned to the Pd−Sn−Pt bond is observed in the $Pd_{0.4}Pt_{0.6}Sn_4$ single crystal (Figure 3a), confirming the synthesis of $Pd_{0.4}Pt_{0.6}Sn_4$ single crystal. Analogously, compared to the Pt−O bond (1.63 Å) in $PtO_2$ ref., the apparent peak located at 2.46 Å can be assigned to Pd−Sn−Pt scattering (Figure 3b), which confirms the presence of $Pd_{0.4}Pt_{0.6}Sn_4$ single crystal and is consistent with the FT-EXAFS pattern of the Pd element.[20] In the FT-EXAFS spectra at the $K$-edge of the Sn element, three characteristic peaks are observed in the $Pd_{0.4}Pt_{0.6}Sn_4$ single crystal (Figure 3c), located at 2.44, 2.46, and 2.55 Å, which are assigned to the bond lengths of Pd−Sn, Pd−Sn−Pt, and Pt−Sn, confirming the synthesis of PdSn$_4$, PtSn$_4$, and Pd$_{0.4}$Pt$_{0.6}$Sn$_4$ single crystals. Subsequent fitting results further confirmed the existence of the Pd−Sn−Pt bond in the Pd$_{0.4}$Pt$_{0.6}$Sn$_4$ single crystal with a bond length of 2.46 Å (Figure S4d−f, Supporting Information), consistent with the computation model after relaxation in density functional theory (DFT) calculations. Meanwhile, for the coordination situations of Pd−Sn, Pt−Sn, and Pd−Sn−Pt coordination, two intensity peaks located at 4.25 and 4.49 Å$^{-1}$ in the *k*-space can be obtained from the corresponding wavelet transformed (WT)-EXAFS analysis profiles (Figure 3d). which further clearly express the coordination of Pd and Pt with Sn elements in *R* space and endorses the semimetal characteristic of Pt, Pd, and Sn atoms in Pd$_{0.4}$Pt$_{0.6}$Sn$_4$ single crystal. Notably, the Pd−Sn−Pt bond length of Pd$_{0.4}$Pt$_{0.6}$Sn$_4$ single crystal is determined to be 2.49 Å, which is longer than that of PdSn$_4$ (2.46 Å), respectively. This phenomenon is attributed to the disruption of the original symmetry of the Pd and Pt centers owing to the formation of Pd−Sn−Pt local units, leading to the stretching of the Pd−Sn and Pt−Sn bonds and ultimately determining the CRR properties. The above findings revealed a change in the microenvironment of the active Pd and Pt centers. The results of XAS analysis reveal the successful formation of the Pd−Sn−Pt local units in the Pd$_{0.4}$Pt$_{0.6}$Sn$_4$ single crystal, and the coordination environment of Pd−Sn−Pt can ensure the full exposure of surface active sites and the stability of the electronic structure. In addition, Figure 3e−g shows highly localized regions and contour plots of the total charge distribution forming covalent bonds between Pd, Pt, and Sn atoms, further implying that the existence of significant charge transfer in the Pd−Sn−Pt local units and that the distribution of electron charges is near the Pd and Pt atoms within the Pd−Sn−Pt local units.[21]

  The charge transfer between Pd, Pt, and Sn atoms in the Pd$_{0.4}$Pt$_{0.6}$Sn$_4$ single crystal is determined by in-situ irradiation X-ray photoelectron spectroscopy (XPS) to elucidate the origin of the electronic structure of the Pd−Sn−Pt local units.[22] Figure S5a,b show the elemental binding energy changes of Pd 3*d* and Pt 4*f* in the Pd$_{0.4}$Pt$_{0.6}$Sn$_4$ single crystal, and it can be found that both characteristic peaks of Pd 3*d* and Pt 4*f* are shifted toward the position of smaller binding energy compared to the "dark-state", which indicates that the electron density of Pd 3*d* has been increased, signifying that the electronic gain of Pd and Pt atoms upon light illumination.[23] Significantly, the binding energies of Sn 3*d* undergo a decrease in light conditions compared to the "dark state" (Figure S5c, Supporting Information), which indicates that the electron density of Pd and Pt atoms increases, and the electron density of the Sn atom decreases in light conditions, indicating that Pd and Pt atoms function as electron acceptors.[12] The XPS spectra of Pt 4*f*, Pd 3*d*, and Sn 3*d* indicate that there is an electronic coupling phenomenon among Pt, Sn, and Pd atoms in the Pd−Sn−Pt local units, and in the $Pd_{0.4}Pt_{0.6}Sn_4$ single crystal, both metallic state and partially oxidized state exist simultaneously.[24] In concert with the XAS results, the in-situ XPS analysis results further provide direct evidence for the Pd−Sn−Pt local electronic structure of $Pd_{0.4}Pt_{0.6}Sn_4$ single crystal.

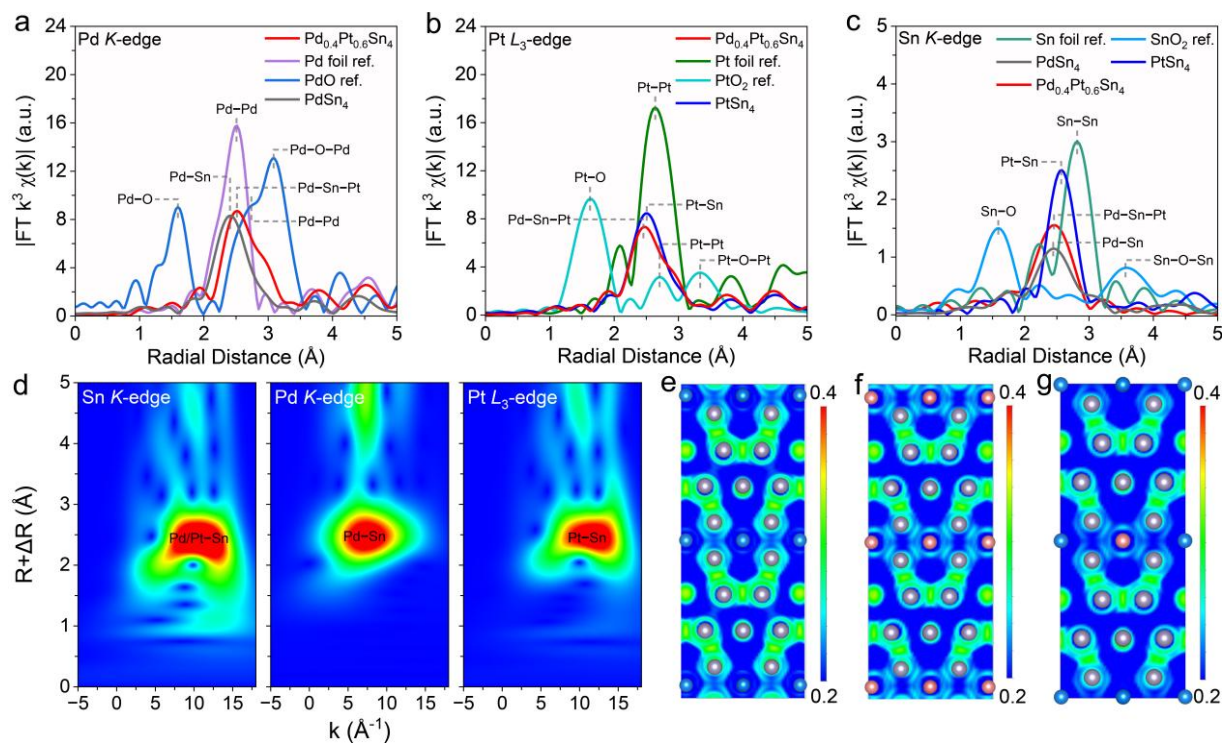

Figure 3. Characterization of Pd, Pt, and Sn chemical state and existing form on the designed photocatalysts. (a) Pd *K*-edge, (b) Pt $L_3$-edge, and (c) Sn *K*-edge EXAFS spectra in *R* space of $Pd_{0.4}Pt_{0.6}Sn_4$ single crystal. (d) WT spectra of Pd *K*-edge, Pt $L_3$-edge, and Sn *K*-edge of the Pd$_{0.4}$Pt$_{0.6}$Sn$_4$ single crystal. The calculated electron localization function diagrams of the (e) PdSn$_4$, (f) PtSn$_4$, and (g) Pd$_{0.4}$Pt$_{0.6}$Sn$_4$ single crystals. Blue, red, and gray spheres represent Pd, Pt, and Sn atoms.

## 2.4 Photochemical Measurements for CRR

To unravel the influence of as-obtained single crystals on the CO$_2$ photoreduction performances, we conducted CO$_2$ photoreduction experiments with acetonitrile (MeCN) aqueous solution and pure water under gas-solid conditions with visible light irradiation ($\lambda \geq$ 420 nm). Moreover, the effects of other factors, such as the ratio of MeCN/H$_2$O and the photocatalyst amount, on photocatalytic activity and C$_2$H$_4$ product selectivity were explored to provide the best photocatalytic conditions. The optimized volume ratio of MeCN:H$_2$O was 4:1 (Figure S6a, Supporting Information) and was adopted to improve CRR catalytic activity and C$_2$H$_4$ selectivity. It was found from Figure S7a that no C$_2$H$_4$, CO, and CH$_4$ were produced without light, a Pd$_{0.4}$Pt$_{0.6}$Sn$_4$ single crystal, and H$_2$O, and H$_2$ was detected in the presence of TEOA and MeCN aqueous solution. Under the optimized conditions, CRR to C$_2$H$_4$ was performed, and the gaseous products in the photocatalytic reduction system were detected by gas chromatography (GC) spectroscopy. C$_2$H$_4$, CH$_4$, and CO products of the Pd$_{0.4}$Pt$_{0.6}$Sn$_4$ single crystal system were detected in the MeCN aqueous solution (Figure S6b, Supporting Information). From the results in Figure 4b, under the same operating conditions, the CO$_2$ conversion rates of PdSn$_4$ and PtSn$_4$ single crystals are only 23% and 25%, respectively, while the CO$_2$ conversion rate of the Pd$_{0.4}$Pt$_{0.6}$Sn$_4$ single crystal is 93%, approximately four times that of PdSn$_4$ and PtSn$_4$ single crystals. Amazingly, as shown in Figure 4a−c, the Pd$_{0.4}$Pt$_{0.6}$Sn$_4$ single crystal exhibits a C$_2$H$_4$ yield rate of 328 μmol·g$^{-1}$·h$^{-1}$, and the selectivity of products and electrons for CO$_2$-to-C$_2$H$_4$ at Pd$_{0.4}$Pt$_{0.6}$Sn$_4$ single crystal reaches as high as 73.1 % and 89.0 % in MeCN aqueous solution, outperforming most previously reported photocatalysts. By contrast, a pure PtSn$_4$ single crystal gives CH$_4$ as the main product at a rate of only 24 μmol·g$^{-1}$·h$^{-1}$, and a PdSn$_4$ single crystal is nearly photo-catalytically incapable toward CO$_2$ reduction to CH$_4$ in MeCN aqueous solution, indicating that the introduction of a Pd site in the PtSn$_4$ single crystal could improve the CRR catalytic activity. The $IQE_{cr}$ of Pd$_{0.4}$Pt$_{0.6}$Sn$_4$ single crystal at 360, 380, 400, 420, 440, 460, 480, 500, and 520 nm were tested and calculated at 2.59 %, 2.41 %, 2.29 %, 2.17 %, 1.47 %, 0.84 %, 0.51 %, 0.90 %, and 0.52 %, respectively (Figure S7b, Supporting Information). To gain further insight into the involvement of surface states, the photocatalytic performances of Pd$_{0.4}$Pt$_{0.6}$Sn$_4$ single crystal and powder were tested in the MeCN aqueous solution system. The main products detected in the Pd$_{0.4}$Pt$_{0.6}$Sn$_4$ powder system were C$_2$H$_4$, CH$_4$, and CO (Figure S8, Supporting Information). Although the specific surface area of Pd$_{0.4}$Pt$_{0.6}$Sn$_4$ powder is higher than that of its single crystal (Figure S9, Supporting Information), the total gas production rate for the Pd$_{0.4}$Pt$_{0.6}$Sn$_4$ single crystal (448 μmol·g$^{-1}$·h$^{-1}$) outperforms that of the Pd$_{0.4}$Pt$_{0.6}$Sn$_4$ powder (336 μmol·g$^{-1}$·h$^{-1}$). This is mainly because the (010) crystal surface exposed by the Pd$_{0.4}$Pt$_{0.6}$Sn$_4$ single crystal has a Dirac nodal arc, which facilitates the adsorption capacity of the catalyst for the CO$_2$ molecule.[7b, 25] At the same time, the exposed (010) crystal surface increases the absorption area of Pd$_{0.4}$Pt$_{0.6}$Sn$_4$ single crystal and enhances the quantum conversion efficiency. In addition, the transient photocurrents and CO$_2$ adsorption isotherms further confirm that the performance of Pd$_{0.4}$Pt$_{0.6}$Sn$_4$ powder was slightly lower than that of its single crystal (Figure S10, Supporting Information).

The photocatalytic performance of Pd$_{0.4}$Pt$_{0.6}$Sn$_4$ single crystal toward CO$_2$ reduction was evaluated under simulated solar irradiation in a pure water system. The Pd$_{0.4}$Pt$_{0.6}$Sn$_4$ single crystal produces CH$_4$ as the main product but at a rate of only 97 μmol·g$^{-1}$·h$^{-1}$ under the pure water conditions, which is about 4.85 and 4.41 times lower than that at PdSn$_4$ and PtSn$_4$ single crystals, respectively (Figure S7c,d, Supporting Information). Additionally, after seven runs of photocatalytic testing, there is negligible decay of CRR photoactivities over the Pd$_{0.4}$Pt$_{0.6}$Sn$_4$ single crystal, as shown in Figure 4d and Figure S7e (Supporting Information), and the Pd$_{0.4}$Pt$_{0.6}$Sn$_4$ single crystal retains 89.3% of its original catalytic activity, indicating excellent stability and durability in the photocatalytic reaction. The morphology and crystal structure are unchanged (Figure S11, Supporting Information). To further determine the carbon source of the resultant products, carbon dioxide carrying the isotopes of $^{13}C$ was used as a reactant, and the products were analyzed by GC-mass spectrometry (GC-MS). The results are shown in Figure S6c (Supporting Information). It is found that the products are $^{13}CO$ (m/z=29), $^{13}CH_4$ (17), and $^{13}C_2H_4$ (30), demonstrating that the photocatalytic product originates from $CO_2$ reduction rather than the decomposition of the catalysts and other carbon-containing species. To the best of our knowledge, the efficiency and selectivity of $Pd_{0.4}Pt_{0.6}Sn_4$ single crystal production in this work outperform most of the reported photocatalysts in MeCN aqueous solution[2, 26] and also rank among the top of the reported $CH_4$ production photocatalysts in non-MeCN medium (Figure 4e).

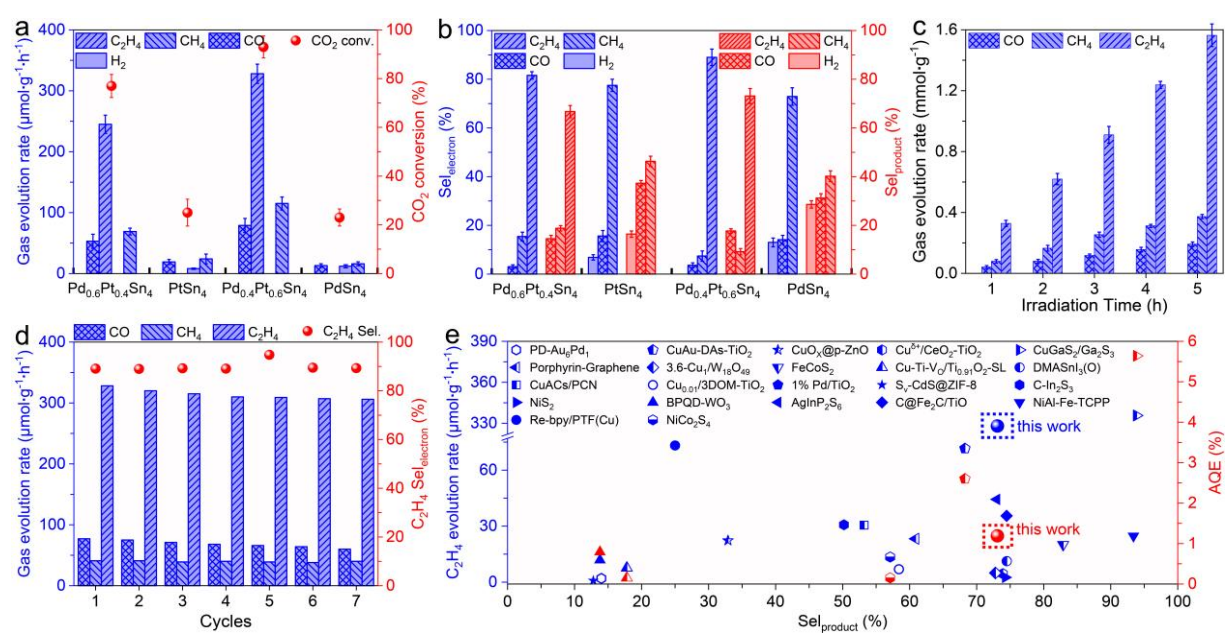

Figure 4. Photocatalytic CRR performances. (a) Product formation rate and $CO_2$ conversion, (b) Electron-based selectivity and product selectivity of $Pd_{0.4}Pt_{0.6}Sn_4$ single crystal in MeCN aqueous solution. (c) Photocatalytic product evolution as a function of light irradiation time on $Pd_{0.4}Pt_{0.6}Sn_4$ single crystal in MeCN aqueous solution. (d) Product formation rate and $C_2H_4$ electron-based selectivity on $Pd_{0.4}Pt_{0.6}Sn_4$ single crystal over seven cycling tests in MeCN

aqueous solution. (e) Comparison of the AQE and produced $C_2H_4$ under Xe lamp over $Pd_{0.4}Pt_{0.6}Sn_4$ single crystal with those of reported catalysts.

## 2.5. Mechanistic Investigation and Theoretical Analysis

To further verify the reaction intermediates, surface carbon, and oxygen species, the intermediates generated over $PdSn_4$ and $Pd_{0.4}Pt_{0.6}Sn_4$ single crystals during CRR were monitored by in-situ electron paramagnetic resonance (EPR) spectra, in-situ near ambient pressure XPS (NAP-XPS), and in-situ diffuse reflectance-infrared Fourier-transform spectroscopy (DRIFTS) measurements. When 5,5-dimethyl-1-pyrroline N-oxide (DMPO) was used as a radical trapper after 10−30 min of irradiation, the characteristic signal of the DMPO-trapped carbon radicals appeared in the $Pd_{0.4}Pt_{0.6}Sn_4$ single crystal system, which might be due to the formation of the primary •$CO_2^-$ radical intermediate. However, •$CO_2^-$ radical intermediate is almost undetectable on the $PdSn_4$ single crystal because the catalytic sites are insufficient to activate $CO_2$ (Figure 5a; Figure S12a, Supporting Information). In-situ EPR spectra analysis indicated that the $Pd_{0.4}Pt_{0.6}Sn_4$ single crystal exhibited a pronounced *H signal in the absence of $CO_2$, which was then significantly attenuated upon the introduction of $CO_2$ (Figure 5b and Figure S12b, Supporting Information). This indicates that the $Pd_{0.4}Pt_{0.6}Sn_4$ single crystal can accelerate the formation of active *H, which is then rapidly consumed during the hydrogenation steps in the CRR process. In contrast, the indigenous active *H formation on $PdSn_4$ single crystal hampers hydrogenation, thereby explaining its low CRR (Figure S12c,d, Supporting Information). This is consistent with the results of evaluating the Gibbs free energy change of $H_2O$ dissociation to generate *H on the $PdSn_4$ and $Pd_{0.4}Pt_{0.6}Sn_4$ models (Figure S12e,f, Supporting Information). As shown in Figure 5d, under ultra-high vacuum conditions, the C1s spectrum shows a distinct peak at 284.5 eV, which is well allocated to the unsteady carbon element on the $Pd_{0.4}Pt_{0.6}Sn_4$ single crystal. The C1s peaks at 284.3, 285.1, 286.1, and 289.3 eV are assigned to C=O, C−C/C−H, C−O, and −COO, and increase with the passage of irradiation time. Furthermore, the O 1$s$ spectrum further confirms that $CO_2$ was reduced to form oxygen-containing species (Figure 5e).

As shown in Figure 5c of $Pd_{0.4}Pt_{0.6}Sn_4$ single crystal and Figure S12g (Supporting Information) of $PdSn_4$ single crystal, new infrared peaks were detected in both samples near 1570 cm$^{-1}$, which is attributed to the COOH* group, a key intermediate for the reduction of $CO_2$ to CO(g) or $CH_4$(g). The absorption bands near 1114 and 1026 cm$^{-1}$ belong to the $CH_3O^*$, and the peak at 1065 cm$^{-1}$ is ascribed to the nonplanar vibration (O=C−H) of CHO*;[27] both the $CH_3O^*$ and CHO* are pivotal intermediates of $CO_2$(g) photoreduction to $CH_4$(g). Moreover, the peaks at 2040 and 2174 cm$^{-1}$ are indexed to the adsorbed and free state CO products, respectively. Besides, the peaks at 1207 and 1306 cm$^{-1}$ are respectively identified as the symmetric and asymmetric stretching vibrations of $HCO_3^-$, while the peak at 1412 cm$^{-1}$ is attributed to the $CO_3^{2-}$ produced during the reaction.[28] The bands at 1507−1570 cm$^{-1}$ could be assigned to the C=O stretching of carbonyl intermediates and asymmetric vibration of *OC−COH (key intermediate for $C_2H_4$), respectively.[26p] Most importantly, the absorption peaks of the *$CH_2$= (896 cm$^{-1}$ for C−H bending vibration) were observed, indicative of a good agreement with the previously reported mechanism of yielding $C_2H_4$.[29] It is reasonably speculated that the most likely reaction pathway for this visible light-driven $CO_2$ reduction can be proposed as in Table S3 (Supporting Information). Concretely speaking, $CO_2$ molecules were adsorbed on the $Pd_{0.4}Pt_{0.6}Sn_4$ single crystal to form the Sn−OCHOH−Pd, while $H_2O$ molecules dissociated into H$^+$ with the help of the generated holes. Then, the Pt−C≡O was produced by a dehydration reaction of *O−CHOH and H$^+$, while the active H$^+$ attacked the charge-rich C atoms in the Pt−C≡O to form the Pt−COH and Sn−OCCO−Pt (Figure 6a). The Sn−OCCOH−Pd was produced by a dehydration reaction of *OC−CHOH and H$^+$, and another H atom in Pt−OHCHCH$_2$−Pt was removed to form *C−CH$_2$ through a similar process, while the H atoms on the Pd species were reduced to $H_2$ molecules by the photogenerated electrons. Finally, the $C_2H_4$ and $H_2$ were desorbed from the surface of the $Pd_{0.4}Pt_{0.6}Sn_4$ single crystal.

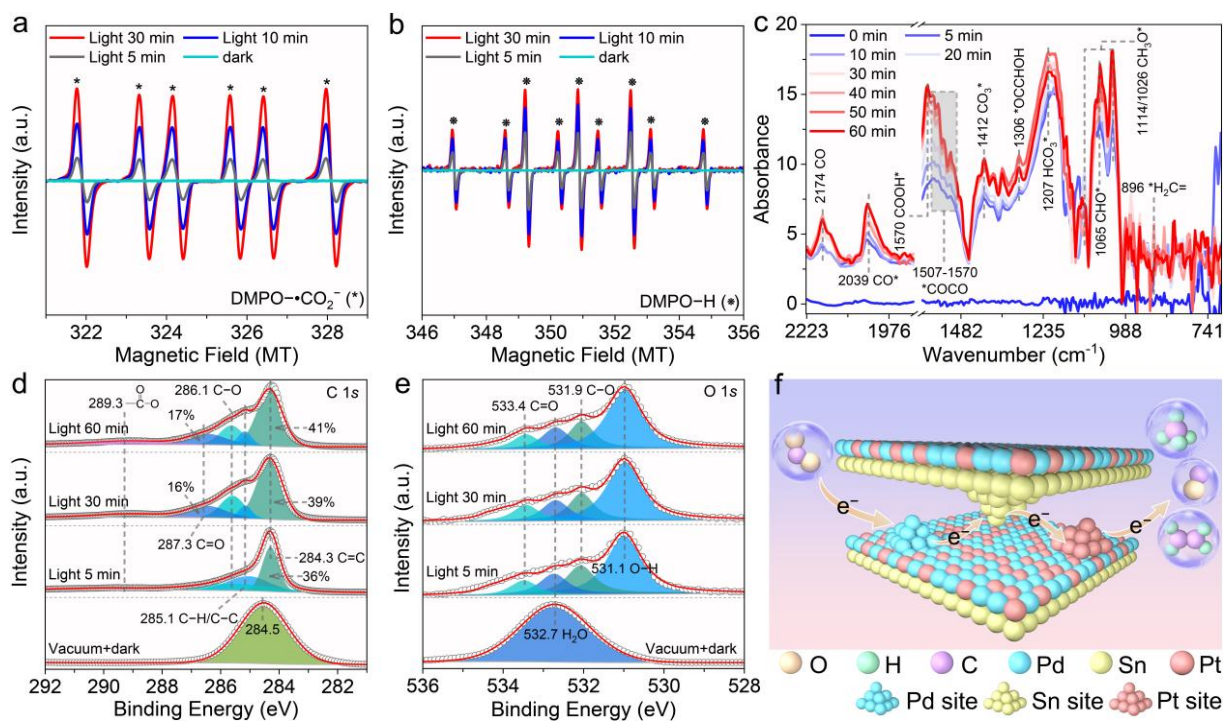

Figure 5. In-situ mechanism study of the photochemical CRR on catalysts. (a) In-situ EPR spectra of $CO_2^{\cdot-}$ adducts of DMPO generated by the photochemical light-driven $CO_2$ conversion over $Pd_{0.4}Pt_{0.6}Sn_4$ single crystal. (b) In-situ EPR spectra of DMPO adducts over $Pd_{0.4}Pt_{0.6}Sn_4$ single crystal in the absence of $CO_2$. (c) In-situ DRIFTS spectra for light-driven $CO_2$ conversion over $Pd_{0.4}Pt_{0.6}Sn_4$ single crystal. In-situ NAP-XPS results of (d) C 1$s$ and (e) O 1$s$ spectra over $Pd_{0.4}Pt_{0.6}Sn_4$ single crystal. (f) Schematic diagram of the Pd−Sn−Pt multi-active sites working in CRR caused by the Pd−Sn−Pt local electronic structure.

Then, in order to deeply study the influence of the local electronic structure of Pd−Sn−Pt on the C−C coupling process, DFT calculations were adopted to illustrate the internal mechanism of $CO_2$-$C_2H_4$. The $PdSn_4$ (010), $PtSn_4$ (010), and $Pd_{0.4}Pt_{0.6}Sn_4$ (010) crystal surface models were built based on our experimental results, and the corresponding Gibbs free energies for *CO, *CO−CO, *OC−COH, and *OC−CHOH adsorptions were calculated (Figure S13 and S14, Supporting Information). The enhanced *OC−COH and *OC−CHOH adsorption on the $Pd_{0.4}Pt_{0.6}Sn_4$ (010) crystal surface can be attributed to the altered electronic structure, as suggested by the bulk band structures and projected density of states (PDOS) calculation (Figure S15, Supporting Information). On PdSn$_4$ and PtSn$_4$ models, the basic steps of reducing CHO* to CH$_3$O* are a highly successive exothermic process. However, the hydrogenation process from CHO* to CH$_3$O* is endothermic, which is energetically less favorable (Figure 6b). The Gibbs free energy profiles of *CO$_2$ reduction to *CO, *CO dimerization to *CO−CO, and *C$_2$H$_4$ formation are shown in Figure 6c and Figure S16 (Supporting Information). The conversion process of CO$_2$-to-C$_2$H$_4$ on the Pd$_{0.4}$Pt$_{0.6}$Sn$_4$ model is more favorable in thermodynamics than that of CH$_4$. Moreover, the desorption energy of CO is 0.42 eV, which is much lower than the hydrogenation energy of *CO to CHO* (4.53 eV), demonstrating that CH$_4$ is not the main product. The coexistence of Pd−Sn−Pt multi-active sites facilitates the production of the *CO as well as the 2*CO → *CO−CO process.[26p] The evolution of C$_2$H$_4$ and CH$_4$ free energy diagrams is summarized and shows that the formation of *CO−CO and CHO* is considered as two potential determining steps (1.42 and 4.53 eV) for the subsequent C−C coupling process (Figure 6c), while the Pd$_{0.4}$Pt$_{0.6}$Sn$_4$ slab can efficiently lower the formation energy of *CO−CO, suggesting that the coexistence of Pd−Sn−Pt multi-active sites can help to optimize the rate-limiting step and accelerate the CO$_2$-to-C$_2$H$_4$ conversion.[30] Noticeably, the different C−C coupling energy barriers of two unsaturated reaction intermediates (*CHOH and *CHO) were evaluated in this work. The free energy of *OC−COH formation was 0.96 eV, which was lower than other coupling pathways (*OC−CHOH was 0.56 eV); nevertheless, the coupling energy barrier value of *OC−CHOH (0.27 eV) was lower than that of *OC−COH (0.63 eV) (Figure 6d), hence the coupling and hydrogenation of *OC−CHOH can generate C$_2$H$_4$. In other words, the coupling of *CO dimerization to generate *OC−CHOH and *OC−COH was the most thermodynamically dominant pathway to form the C−C bond.[31] As for the next step, the *OC−CHOH can be further dehydrated to form *OC−CH; the formation of *OC−CH is $\Delta G = -0.29$ eV. Above all, the combined DFT and empirical results suggest that Pd$_{0.4}$Pt$_{0.6}$Sn$_4$ working in conjunction with Pd−Sn−Pt local electronic structure influences C$_2$H$_4$ selectivity in addition to promoting C−C coupling.

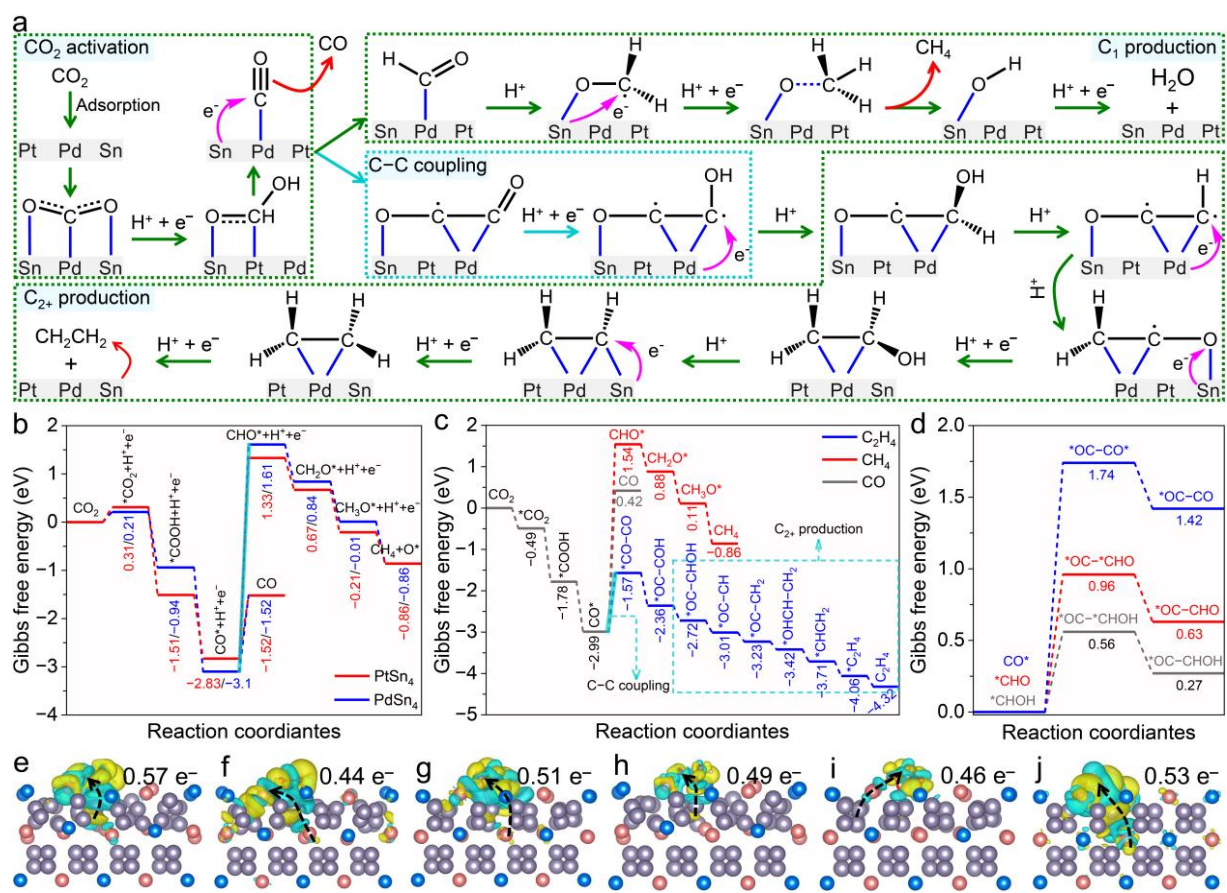

Figure 6. DFT calculation analyzing energy changes. (a) Proposed reaction mechanism for CRR to $C_{2+}$ products on the surface of $Pd_{0.4}Pt_{0.6}Sn_4$ single crystal. (b) The Gibbs free energy diagrams for CRR to CO and $CH_4$ on the surface of $PdSn_4$ and $PtSn_4$ models. (c) The Gibbs free energy diagrams for CRR to $C_2H_4$ and $CH_4$ on the surface of the $Pd_{0.4}Pt_{0.6}Sn_4$ model. (d) Free energy of different intermediates on the surface of the $Pd_{0.4}Pt_{0.6}Sn_4$ model. 3D electron density difference distributions for adsorption of (e) *COOH, (f) *CO, (g) *CO−CO, (h) *OC−COH, (i) *OC−CHOH, and (j) *$C_2H_4$ on the optimized $Pd_{0.4}Pt_{0.6}Sn_4$ model surface (cyan and yellow regions represent electron depletion and accumulation).

To further explore how the multi-active sites generated by the Pd−Sn−Pt local electronic structure reduce the energy barrier of C−C coupling of *CO (Figure 5i), charge density difference analysis was performed on adsorbed *COOH, *CO, *CO−CO, *OC−COH, *OC−CHOH, and *$C_2H_4$ (Figure 6e−j). As shown in Figure 6f, the charge distribution on Pd and Pt atoms in the $Pd_{0.4}Pt_{0.6}Sn_4$ model is asymmetric, which weakens the electrostatic repulsion between CO*, thereby promoting the dimerization of *CO into *CO−CO (Figure 6g).[26p] It can be found that the C−C distance of the *OC−CHOH molecular structure is approximately 1.34 Å, while the ∠C−C site is about 116° (Figure 6i). The distance between the two active sites that form *OC−CHOH is only 2.6 Å, indicating that an Sn−O bond is formed during the absorption of *OC−CHOH on the surface of the $Pd_{0.4}Pt_{0.6}Sn_4$ model (Figure 6a).[32] That is to say, after the introduction of Pt atoms into $PdSn_4$ models, the Pd−Sn dual sites turned out to be the Pd−Sn−Pt multi-active sites, which enabled the tight adsorption of *OC−CHOH, thus helping to boost the C−C coupling step.[2] In addition, the Pd and Pt tend to lose electrons, and the charges tend to aggregate around the Sn atoms, leading to the higher charge density of the Sn−O bond between $Pd_{0.4}Pt_{0.6}Sn_4$ models and the *OC−CHOH (Figure 6i). According to Bader charge analysis, the *OC−COH in the Sn−OCCOH−Pd structure gains 0.49 $e^-$ from the Sn atoms of the Pd−Sn−Pt local electronic structure (Figure 6h). In short, the $Pd_{0.4}Pt_{0.6}Sn_4$ model leads to higher charge density around the introduced Sn atoms, which is beneficial to establishing the Pd−Sn−Pt multi-active sites for compactly bonding with the *OC−CHOH, thus promoting C−C coupling to selectively produce $C_2H_4$. The electron localization and electrostatic potential for adsorption of *COOH, *CO, and *COCO on the surface of $Pd_{0.4}Pt_{0.6}Sn_4$ models are shown in Figure S17 (Supporting Information), where Pt atoms can transfer more charges to the neighboring Sn atoms than Pd atoms, showing stronger bonding interactions to reduce the activation energy of *CO coupling, thereby promoting the formation of $C_2H_4$. This results in the intermediates generated during the CRR process being more likely to bind to the Pt atoms rather than the Pd atoms of Pd−Sn−Pt local units.[2] The theoretical calculations are in excellent agreement with the experimental results, indicating that Pd−Sn−Pt local electronic structure induces more charges to aggregate around Pd and Pt atoms, can optimize the electronic structure of $Pd_{0.4}Pt_{0.6}Sn_4$ models, resulting in its lowest reaction barrier for *OC−COH and *OC−CHOH. Therefore, both experimental data and theoretical calculations have demonstrated that the Pd−Sn−Pt multi-active site effectively enhances the performance of $CO_2$ reduction to $C_2H_4$.

## 3. Conclusion

We developed a series of $Pd_{1-x}Pt_xSn_4$ single crystals, proving to be an effective photocatalyst for the reduction of $CO_2$ to $C_2H_4$. Both experimental observations and DFT calculations corroborate that Pd−Sn−Pt local electronic structure induces more charges to aggregate around Pd and Pt atoms, resulting in Pd−Sn−Pt multi-active sites. Mechanistic analysis reveals that the Pd−Sn−Pt multi-active sites strengthen the interaction between the *OC−COH and *OC−CHOH, a stark contrast to what is observed in the $PdSn_4$ single crystal. Accordingly, the $C_2H_4$ generation rate, $C_2H_4$ selectivity, and $C_2H_4$ electron-based selectivity of the $Pd_{0.4}Pt_{0.6}Sn_4$ single crystal are 328 $\mu mol \cdot g^{-1} \cdot h^{-1}$, 73.1%, and 89%, respectively. This study experimentally/theoretically verifies that the multi-active sites constructed by atomic doping can effectively reduce the energy barrier of C−C coupling, thereby enhancing the selectivity of $CO_2$-to-$C_2H_4$ and paving a new path for obtaining the performance of efficient $CO_2$ photoreduction $C_{2+}$ products.

## 4. Experimental Section

**Materials and Reagents**

Palladium (Pd, purity 99.99%), platinum (Pt, purity 99.99%), and tin (Sn, purity 99.99%) powders were all obtained from Shanghai Aladdin Biochemical Technology Co., Ltd in China. The chemical reagents used in this experiment were all of analytical purity level and were directly used without any purification treatment.

**Preparation of $Pd_xPt_{1-x}Sn_4$ Single Crystals**

Here, high-purity Pd, Pt, and Sn powders were weighed in stoichiometric proportions, thoroughly ground under an inert atmosphere, and subsequently vacuum-sealed within a quartz ampoule. The ampoule was heated to 950 °C at a rate of 2 °C·min$^{-1}$ and held at this elevated temperature for 48 h. Following this, the melted sample was cooled to 310 °C at a cooling rate of 10 °C·min$^{-1}$, annealed for 6 h, then further cooled to 240 °C at a slower rate of 1 °C·min$^{-1}$ for a 48 h phase stabilization period.[25c, 33] Finally, the sample naturally cooled down to room temperature. The heating procedures for the synthesis of single crystals in different proportions are all the same. In this work, unless otherwise specified, all characterizations were performed using single-crystal wafers. The $Pd_{0.4}Pt_{0.6}Sn_4$ polycrystalline powder, exhibiting a black color, was prepared by grinding and subsequent ball milling of the initially bright silver $Pd_{0.4}Pt_{0.6}Sn_4$ single crystal.

**Materials Characterizations**

TEM imaging and the corresponding EDS mapping were performed on a JEOL JEM 2100F FEG instrument operated at 200 kV, with an Oxford Instruments ULTIM silicon drift detector used for EDS data acquisition. The atomic-resolution HADDF-STEM images were obtained using a double Cs-corrected Thermo Scientific microscope at 300 kV. The PXRD patterns were recorded on a Rigaku Miniflex-600 diffractometer with Cu Kα radiation (λ = 1.5418 Å), operated at 40 kV voltage and 30 mA current, over a 2$\theta$ range of 10° to 70°. The specific surface areas of both single crystals and their powders were determined via nitrogen adsorption measurements using an ASAP 2460 (U.S.A., Micromeritics). The UV-vis DRS was acquired on the Shimadzu UV-2550 instrument. The PL spectra and TRPL decay curves were collected at room temperature on a FLS980 spectrophotometer, employing an excitation wavelength of 320 nm. The electrochemical properties, including the EIS and transient photocurrent response, were evaluated on a CHI 760E electrochemical workstation equipped with a standard three-electrode system (a saturated Ag/AgCl reference electrode, a Pt mesh counter electrode, and a working electrode prepared by coating the single crystals onto carbon paper). The Hall effect measurements were implemented in a Quantum Design Physical Property Measurement System (QD-PPMS).

**Supporting Information**

Supporting Information is available from the Wiley Online Library or the author.

**Acknowledgements**

This work is supported by the National Natural Science Foundation of China (12274471, and 12404165), Guangdong Basic and Applied Basic Research Foundation (Grant No. 2025A1515010311), Guangdong Major Project of Basic Research (2025B0303000004), Guangzhou Science and Technology Programme (No. 2024A04J6415), the Open Research Fund of State Key Laboratory of Quantum Functional Materials (QFM2025KF004), the State Key Laboratory of Optoelectronic Materials and Technologies (Sun Yat-Sen University, No. OEMT-2024-ZRC-02), Key Laboratory of Magnetoelectric Physics and Devices of Guangdong Province (Grant No. 2022B1212010008), and Research Center for Magnetoelectric Physics of Guangdong Province (2024B0303390001). Lingyong Zeng acknowledges the Postdoctoral Fellowship Program of CPSF (GZC20233299) and the Fundamental Research Funds for the Central Universities, Sun Yat-sen University (24qupy092).

**Conflict of Interest**

The authors declare no conflict of interest.

**Data Availability Statement**

The data that support the findings of this study are available from the corresponding author upon reasonable request. In this work, the tests were performed using single-crystal wafers unless otherwise specified.

**Keywords**

Dirac nodal arc semimetal, multi-active sites, $CO_2$ photoreduction, $C_{2+}$ products

# Supporting Information

**Efficient Photocatalytic $CO_2$ Reduction to $C_{2+}$ Products with $Pt_{1-x}Pd_xSn_4$ Dirac Nodal Arc Semimetal**

*Kangwang Wang, Jie Zhan, Jun Liu, Zaichen Xiang, Wanyi Zhang, Lingyong Zeng, Kai Yan\*, Yan Sun\*, and Huixia Luo\**

K. Wang, J. Liu, Z. Xiang, W. Zhang, L. Zeng, Prof. H. Luo
School of Materials Science and Engineering, State Key Laboratory of Optoelectronic Materials and Technologies, Guangdong Provincial Key Laboratory of Magnetoelectric Physics and Devices, Key Lab of Polymer Composite & Functional Materials,
Sun Yat-sen University,
Guangzhou 510275, China
E-mail: luohx7@mail.sysu.edu.cn

Prof. K. Yan
School of Environmental Science and Engineering,
Sun Yat-sen University,
Guangzhou 510275, China
E-mail: yank9@mail.sysu.edu.cn

J. Zhan, Prof. Y. Sun
Shenyang National Laboratory for Materials Science, Institute of Metal Research,
Chinese Academy of Sciences,
Shenyang 110016, China
School of Materials Science and Engineering,
University of Science and Technology of China, Shenyang 110016, China
E-mail: sunyan@imr.ac.cn

## Section I: Experiment section

**Characterization**

**X-ray Absorption Spectra measurements:** XANES and EXAFS measurements of Pd, Pt, and Sn were collected at the beamline BL14W1 of the Shanghai Synchrotron Radiation Facility (SSRF, Shanghai) in a fluorescence mode at room temperature. Data reduction, data analysis, and EXAFS fitting were performed and analyzed using the Athena and Artemis programs of the Demeter data analysis packages, which utilize the FEFF6 program to fit the EXAFS data. The energy calibration of the sample was conducted using a standard Pd foil, Pt foil, and Sn foil, which were simultaneously measured as a reference. A linear function was subtracted from the pre-edge region, then the edge jump was normalized using Athena software. WT spectra were also employed using the software package developed by Funke and Chukalina using the Morlet wavelet with $\kappa = 20$, $\sigma = 1$.

**In-situ XPS measurements:** The in-situ XPS tests were carried out on the Thermo Scientific K-Alpha system with monochromatic Al-K$\alpha$ radiation ($h\upsilon = 1486.6$ eV), operated at 12 kV. The in-situ XPS spectra were recorded before irradiation, after irradiation for 15 and 30 min, and without irradiation for 15 min by a Xe lamp (300 W Xe lamp, Beijing Perfect Light Technology Co., Ltd, PLS-SXE300D).

**In-situ DRIFTS measurements:** The in-situ DRIFTS measurements were conducted by a Thermo Scientific Nicolet iS50 spectrometer with a liquid nitrogen-cooled mercury-cadmium-telluride (MCT) detector in the system. In a typical procedure, the semimetal was placed in the chamber before sealing and then purging with Ar gas for 60 min. Typical signals of several intermediates were captured after the introduction of a flowing $CO_2(g)$ and $H_2O(g)$ ($CO_2/H_2O = 8/1$) under dark conditions for 10 min and then visible light irradiation for 0−60 min. Another test was performed with a flowing $CO_2(g)$ under dark conditions for 10 min and then visible light irradiation for 0−60 min.

**In-situ NAP-XPS measurements:** The in-situ NAP-XPS measurements were obtained on the beamline BL02B1 of SSRF equipped with a Xe lamp (300 W Xe lamp, Beijing Perfect Light Technology Co., Ltd, PLS-SXE300D) as the illumination source.

**In-situ EPR measurements:** The in-situ EPR measurements were obtained on the Bruker model JEOL JES-FA200 spectrometer equipped with a Xe lamp (300 W Xe lamp, Beijing Perfect Light Technology Co., Ltd, PLS-SXE300D) as the illumination source, with 5,5-dimethyl-1-pyrroline N-oxide (DMPO) as a radical trapper.

**$CO_2$-TPD measurements:** The $CO_2$ temperature-programmed desorption measurement ($CO_2$-TPD, AUTO CHEM 2920) was used to confirm the active adsorption sites of the catalysts.

**Apparent quantum efficiency (*AQE*):** The *AQE* was determined under the above photocatalytic reaction conditions, except that the Xe lamp was used as the irradiation source. The wavelength-dependent *AQE* was measured under the same photocatalytic reaction conditions, except for the monochromatic light wavelengths (360, 380, 400, 420, 440, 460, 480, 500, and 520 nm). At the same time, the *AQE* and corresponding *IQE* ($IQE_{cr}$) were roughly calculated from the equation:[1]

$$AQE = \frac{N_{electrons}}{N_{photons}} \times 100\%$$  Equation S1

$$N_{electrons} = n_{electrons} \times N_A$$

Where $N_A$ is the Avogadro constant ($6.02 \times 10^{23}$ mol$^{-1}$).

$$N_{photons} = \frac{ItS_1}{Q}; Q = \frac{hc}{\lambda}; I = \frac{P}{S_2}$$  Equation S2

Where *I*, $S_1$, *t*, and *Q* represent light intensity (W·m$^{-2}$), irradiation area (in this work, $15.89 \times 10^{-4}$ m$^2$), irradiation time (s), and photon energy (J), respectively. *P* and $S_2$ are the light power (W) and aperture area (in this work, $3.14 \times 10^{-4}$ m$^2$) of the light power meter, respectively.[2]

$$N_{photons} = \frac{\frac{P}{S_2}tS_1}{\frac{hc}{\lambda}} = \frac{PtS_1\lambda}{S_2hc} = \frac{PS_1}{S_2}\frac{\lambda t}{hc} = E\frac{\lambda t}{hc} \qquad \text{Equation S3}$$

Where $h$, $c$, and $\lambda$ represent Planck's constant (6.626×10$^{-34}$ j·s), light speed (3×10$^{-34}$ m·s$^{-1}$), and monochromatic light wavelength, respectively.

$$IQE_{cr} = \frac{AQE}{A} \times 100\% \qquad \text{Equation S4}$$

Where A is the absorption of the sample.[3]

The values of $E$ at different wavelengths ($\lambda$ = 360, 380, 400, 420, 440, 460, 480, 500, and 520 nm) were calculated to be 1.74, 1.75, 1.71, 1.67, 1.96, 2.12, 2.33, 0.70, and 0.63 W by Xe lamp (300 W Xe lamp, Beijing Perfect Light Technology Co. Ltd, PLS-SXE300D) source. Taking Pd$_{0.4}$Pt$_{0.6}$Sn$_4$ single crystal at $\lambda$ = 420 nm as an example:

$$AQE_{(C_2H_4)} = \frac{12 \times 16.4 \times 10^{-6} \times 6.02 \times 10^{23}}{1.27 \times 10^{22}} \times 100\% = 0.93\%$$

$$AQE_{(CH_4)} = \frac{8 \times 5.8 \times 10^{-6} \times 6.02 \times 10^{23}}{1.27 \times 10^{22}} \times 100\% = 0.22\%$$

$$AQE_{(CO)} = \frac{2 \times 4.0 \times 10^{-6} \times 6.02 \times 10^{23}}{1.27 \times 10^{22}} \times 100\% = 0.04\%$$

$$Total\ AQE = AQE_{(C_2H_4)} + AQE_{(CH_4)} + AQE_{(CO)} = 1.19\%$$

$$IQE_{cr} = \frac{Total\ AQE}{A} \times 100\% = \frac{1.19\%}{0.5473} \times 100\% = 2.17\%$$

**Photocatalytic CRR measurements:** The photocatalytic CRR measurement was conducted in an online trace gas analysis system with a gas chromatography-mass spectrometry (GC-MS, Agilent GC/MS-7000D) and $^1$H nuclear magnetic resonance ($^1$H NMR) spectroscopy, where 3 mL H$_2$O was injected into the bottom of the photocatalytic reactor before sealing the system, which ensured the H$_2$O vapor was saturated during photocatalysis. In detail, 100 mg of photocatalysts, 20 mL of triethanolamine (TEOA) solution, and 100 mL of acetonitrile (MeCN) aqueous solution (MeCN:H$_2$O = 4:1) were added to the Pyrex glass reaction cell with sonication.

The reactor system was then filled with pure $CO_2$ gas following complete evacuation. After adsorption equilibrium in a dark environment, the reactor system was put under a simulated light source, which was provided by a Xe lamp (300 W Xe lamp, Beijing Perfect Light Technology Co., Ltd, PLS-SXE300D, 100 mW cm$^{-2}$). In addition, the reaction system was controlled at 298 K by circulating water. Finally, the produced gas was analyzed every 1 h through GC and $^1$H NMR spectroscopy (Bruker AVIII HD 600), respectively. The error bars for gas evolution uncertainty represent one standard deviation based on 3 independent samples.

**First-principles Calculations:** The Vienna ab initio simulation package (VASP) was used for density functional theory (DFT) calculations. The revised Perdew-Burke-Ernzerhof (PBE) functional was employed with the basic projector-augmented wave (PAW) method, using a cutoff energy of 420 eV. The bulk lattice structure was relaxed by a k-grid of 3×3×2 for the doped case of $Pd_xPt_{1-x}Sn_4$, and 4×4×3 for the pristine case of $PdSn_4$ and $PtSn_4$, and the (010) surface was built first to build the activity trend for CRR. The bulk electronic band structures and projected density of states (PDOS) are calculated for the simple case, as well as including the spin-orbit coupling (SOC). For the SOC band structures, full relativistic corrected pseudo-potentials are used from the PSEUDOJOJO library. The calculation parameters are as follows: the total energy convergence threshold is set to $1.36 \times 10^{-4}$ eV/atom. The $\Delta G$ was defined as:

$$\Delta G = \Delta E + \Delta E_{\text{ZPE}} - T\Delta S \qquad \text{Equation S5}$$

where $\Delta E$ is the energy difference between the reactants and product obtained through DFT calculations. $\Delta E_{\text{ZPE}}$ and $\Delta S$ are the changes in the zero-point energies (ZPE) and entropy. $T$ represents the temperature and was set as 298.15 K.

## Section II: Figures and Tables

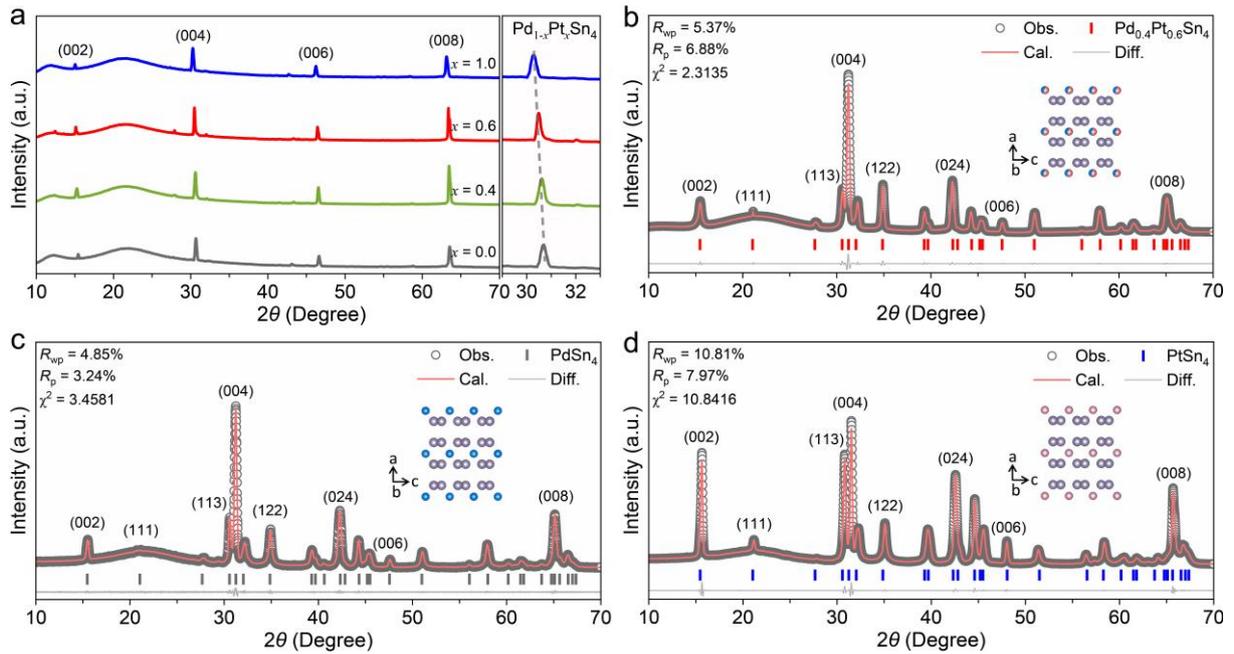

**Figure S1.** (a) XRD pattern taken on a crystal flake of the $Pd_{1-x}Pt_xSn_4$ single crystals (the insets of (a) show the photos of the $Pd_{1-x}Pt_xSn_4$ single crystals, presenting bright silver color). Rietveld refined the PXRD patterns of the (b) $Pd_{0.4}Pt_{0.6}Sn_4$, (c) $PdSn_4$, and (d) $PtSn_4$ powders (the insets show the VESTA drawn unit cell).

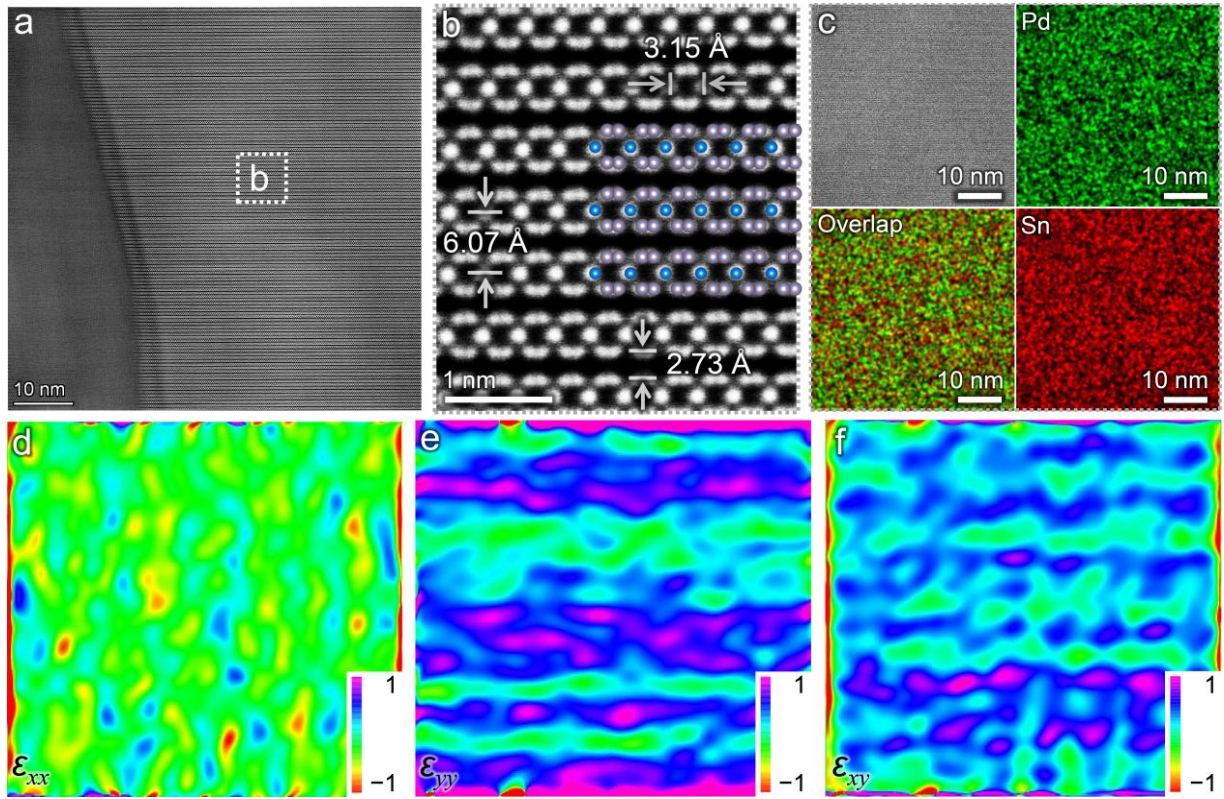

**Figure S2.** (a,b) HAADF-STEM images taken from the [010] direction (the inset of (b) shows the VESTA drawn unit cell), and (c) EDS mapping of PdSn$_4$ single crystal. (d−f) GPA analysis based on the HAADF image (b).

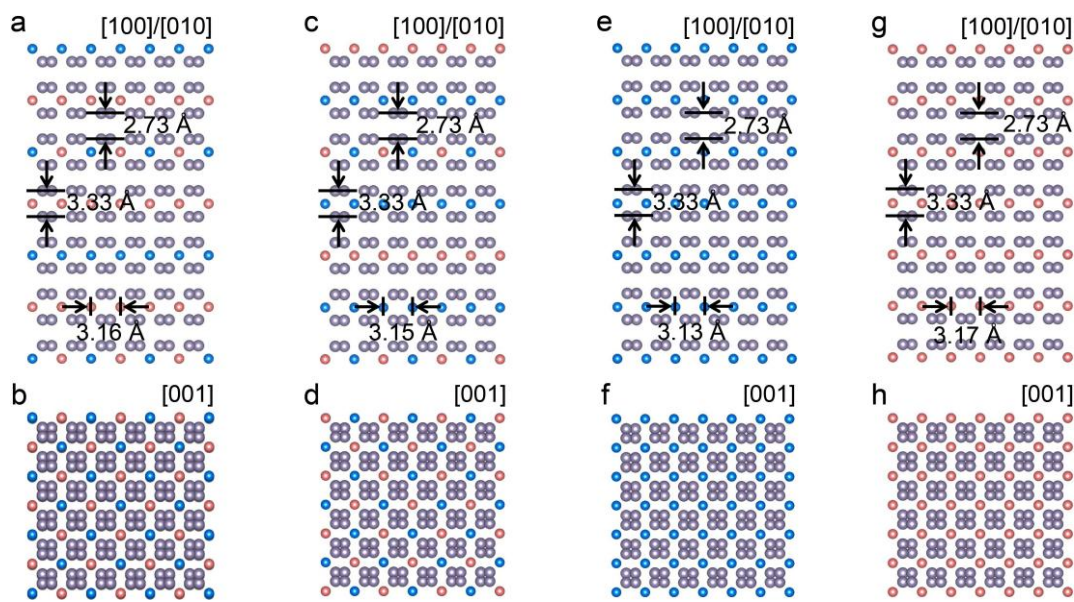

**Figure S3.** Rietveld refined crystal structures of the (a,b) Pd$_{0.4}$Pt$_{0.6}$Sn$_4$, (c,d) Pd$_{0.6}$Pt$_{0.4}$Sn$_4$, (e,f) PdSn$_4$, and (g,h) PtSn$_4$ single crystals, viewed from (a) [100]/[010] and (b) [001] directions, respectively. The gray, blue, and red spheres represent Sn, Pd, and Pt atoms, respectively.

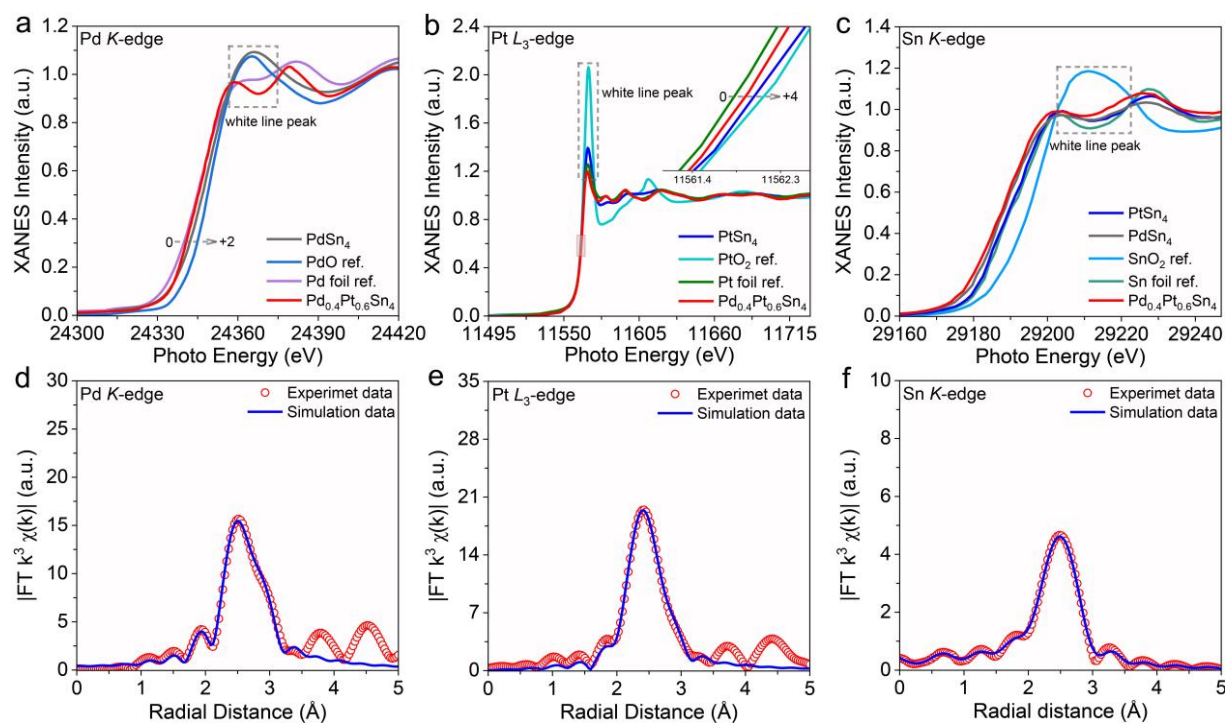

**Figure S4.** (a) Pd *K*-edge, (b) Pt $L_3$-edge, and (c) Sn *K*-edge XANES spectra of the PdSn$_4$, PtSn$_4$, and Pd$_{0.4}$Pt$_{0.6}$Sn$_4$ single crystals. Fitting results of (d) Pd *K*-edge, (e) Pt $L_3$-edge, and (f) Sn *K*-edge in *R*-space for the Pd$_{0.4}$Pt$_{0.6}$Sn$_4$ single crystal.

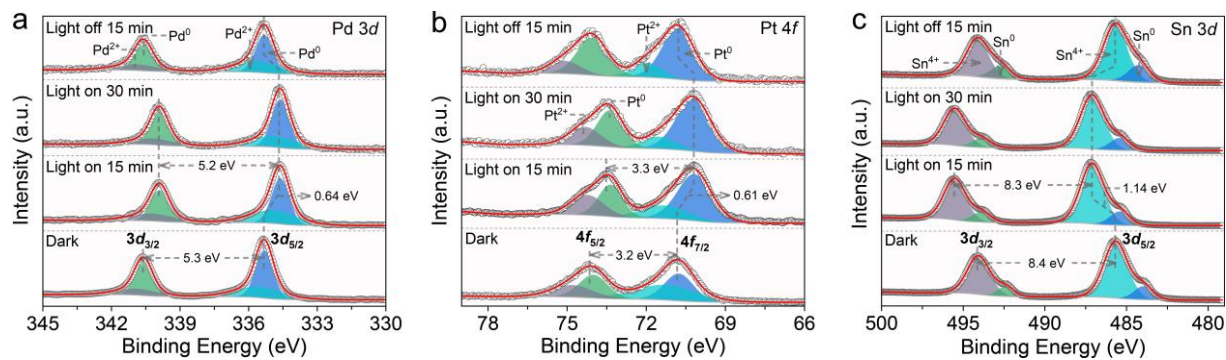

**Figure S5.** In-situ XPS (a) Pd 3*d*, (b) Pt 4*f*, and (c) Sn 3*d* spectra of the Pd$_{0.4}$Pt$_{0.6}$Sn$_4$ single crystal.

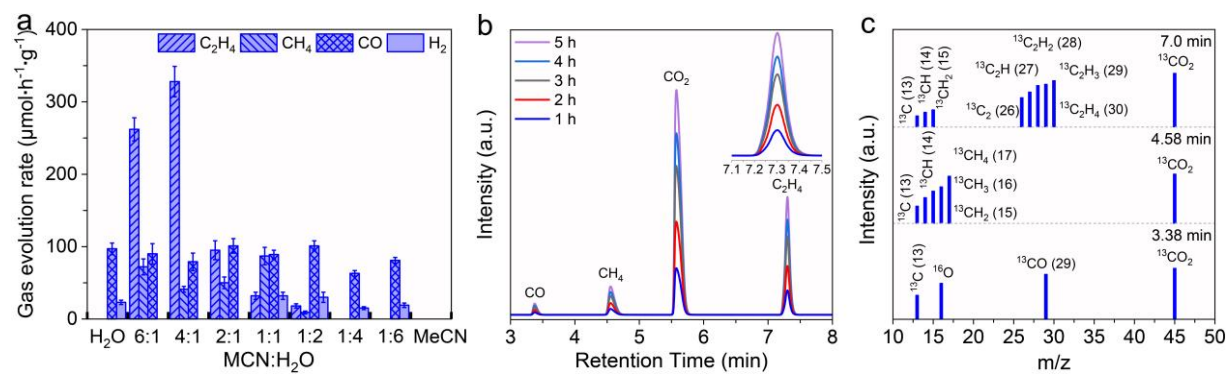

**Figure S6.** (a) Effect of the MeCN/H$_2$O volume ratio on the photocatalytic CRR in the Pd$_{0.4}$Pt$_{0.6}$Sn$_4$ single crystal system. (b,c) GC-MS profiles of CO, CH$_4$, CO$_2$, and C$_2$H$_4$ of the Pd$_{0.4}$Pt$_{0.6}$Sn$_4$ single crystal in MeCN aqueous solution.

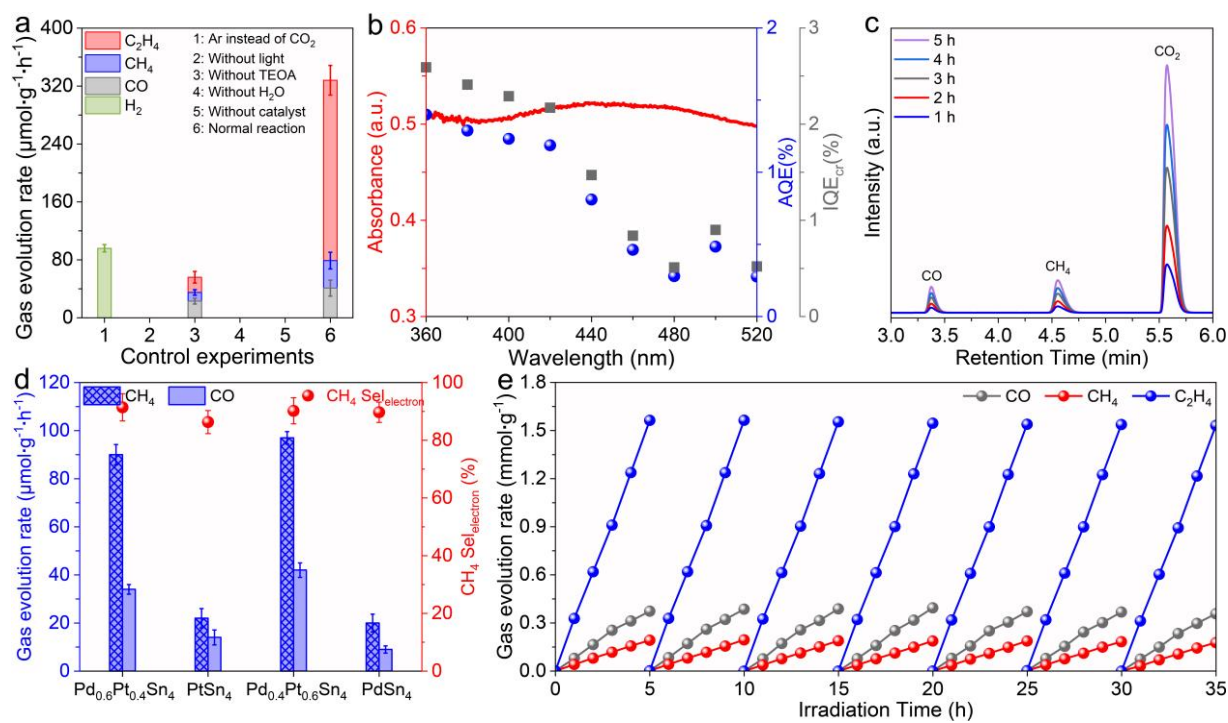

**Figure S7.** (a) Control experiments of catalytic conditions. (b) The AQE, IQE$_{cr}$, and absorption spectrum of the Pd$_{0.4}$Pt$_{0.6}$Sn$_4$ single crystal. (c) GC profile of the gaseous products from CO$_2$ photoreduction using the PdSn$_4$ single crystal in pure water. (d) Product formation rate and CH$_4$ electron-based selectivity of the Pd$_{0.4}$Pt$_{0.6}$Sn$_4$ single crystal in pure water. (e) Long-term photocatalytic stability test of the Pd$_{0.4}$Pt$_{0.6}$Sn$_4$ single crystal.

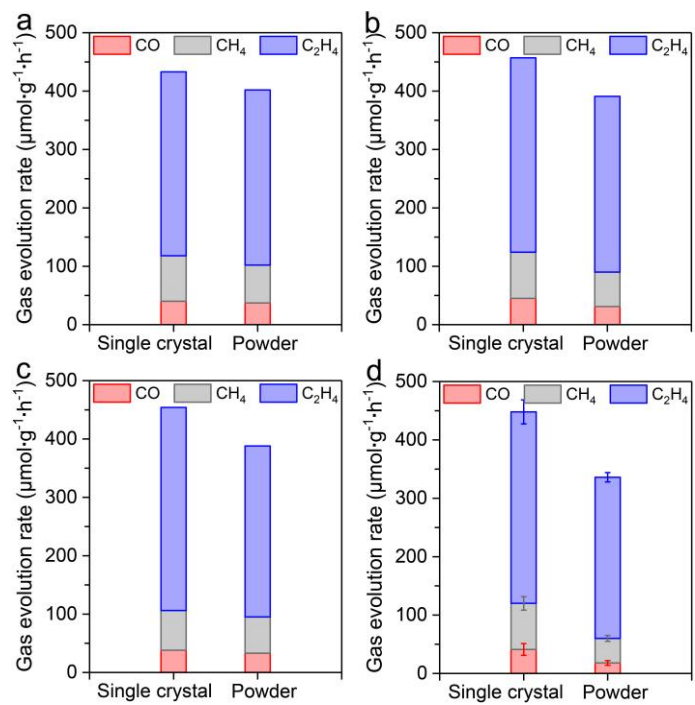

**Figure S8.** Product formation rate of Pd$_{0.4}$Pt$_{0.6}$Sn$_4$ single crystal and powder in MeCN aqueous solution: (a) first test, (b) second test, (c) third test. (d) Product formation rate of Pd$_{0.4}$Pt$_{0.6}$Sn$_4$

single crystal and powder in MeCN aqueous solution (The error bars for gas evolution uncertainty represent one standard deviation based on 3 independent samples.).

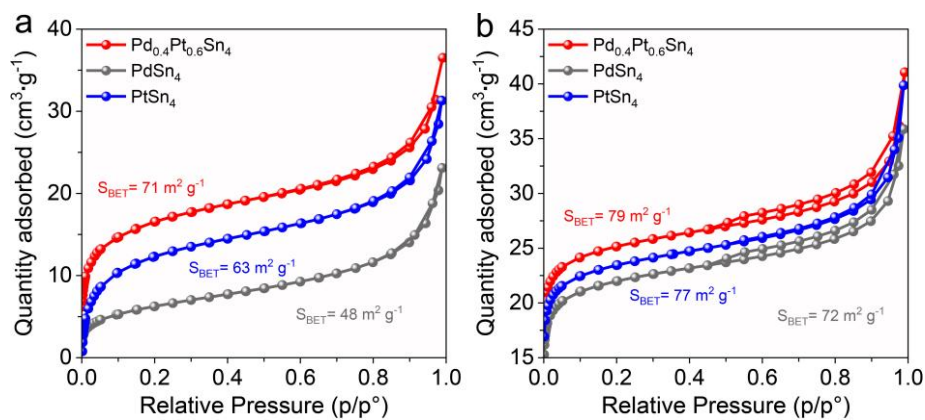

**Figure S9.** Nitrogen adsorption-desorption isotherms of the $PdSn_4$, $PtSn_4$, and $Pd_{0.4}Pt_{0.6}Sn_4$ (a) single crystals and (b) powders, respectively.

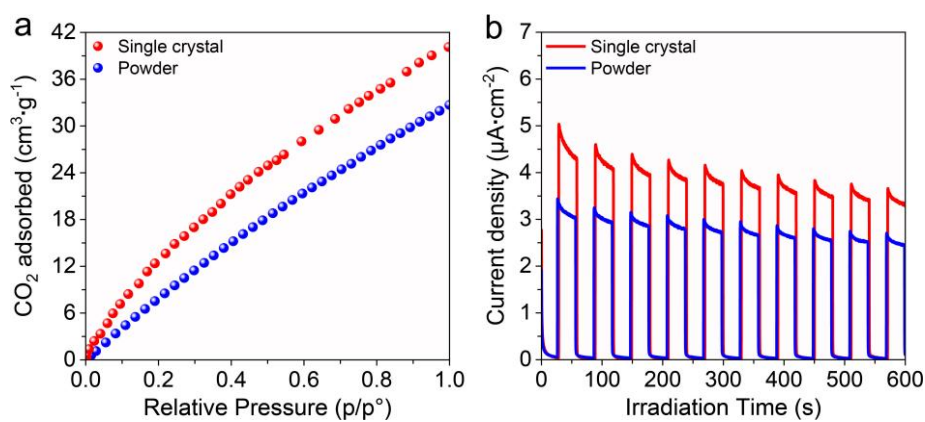

**Figure S10.** (a) $CO_2$ adsorption isotherms (298 K) of the $Pd_{0.4}Pt_{0.6}Sn_4$ single crystal and powder. (b) Transient photocurrents (1.0 M KOH) spectra of the $Pd_{0.4}Pt_{0.6}Sn_4$ single crystal and powder.

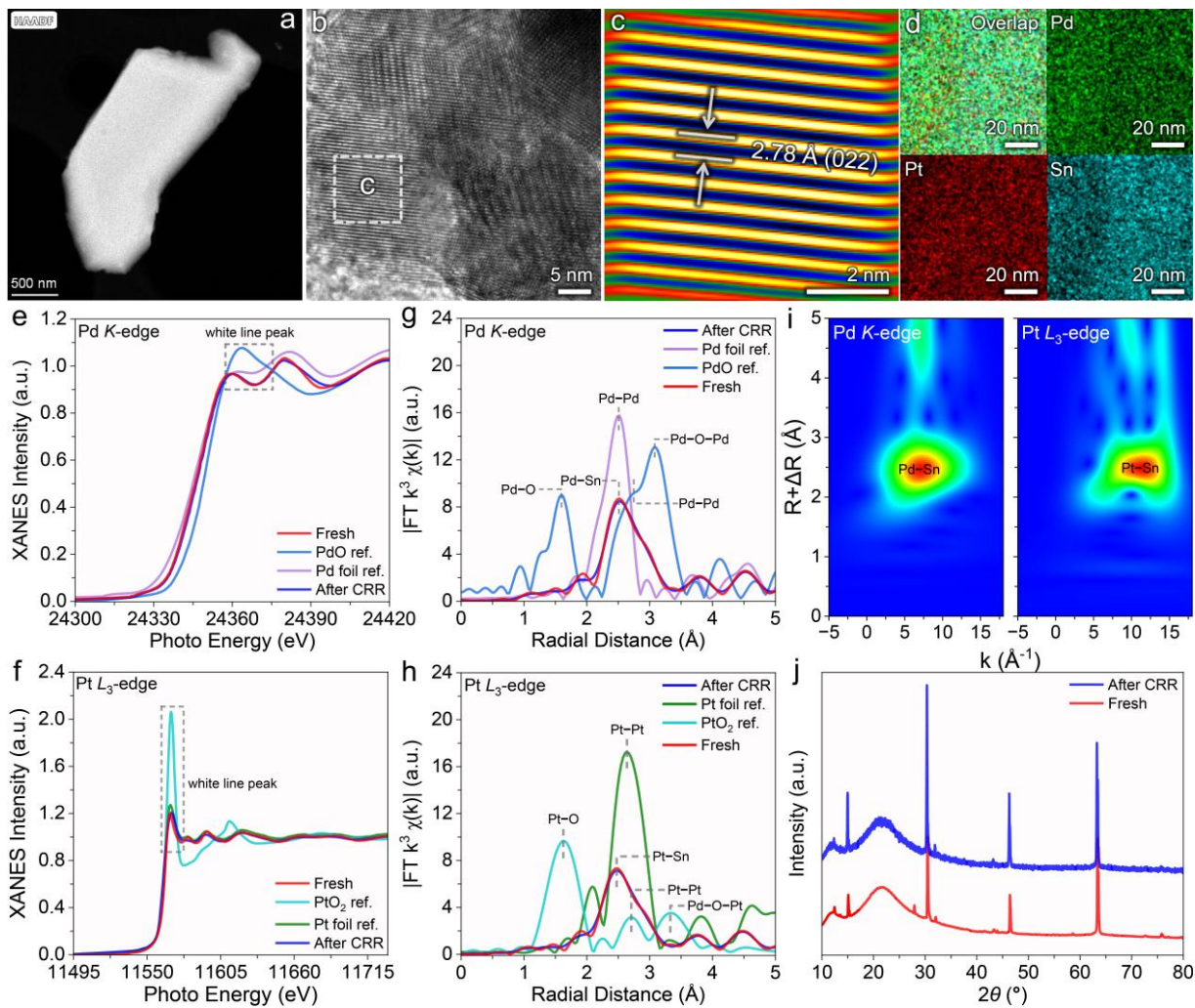

**Figure S11.** (a−c) TEM images, (d) EDS mapping, (e) Pd *K*-edge and (f) Pt *L*$_3$-edge XANES spectra, (g) Pd *K*-edge and (h) Pt *L*$_3$-edge EXAFS spectra, (i) WT of Pd *K*-edge and Pt *L*$_3$-edge, and (j) XRD pattern of the Pd$_{0.4}$Pt$_{0.6}$Sn$_4$ single crystal after CRR (seven cycles).

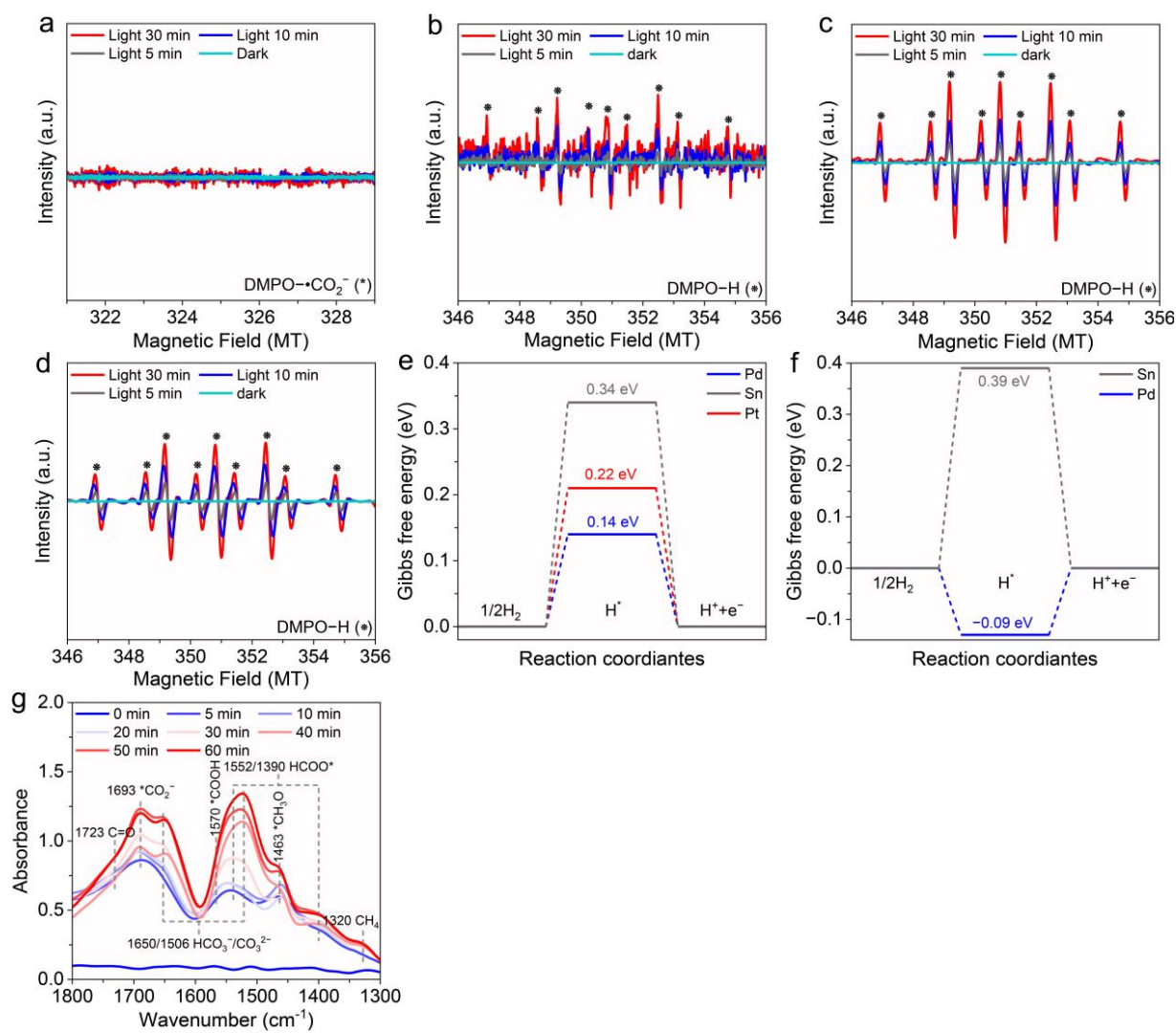

**Figure S12.** (a) EPR spectra of $CO_2^{\bullet-}$ adduct of DMPO generated by the photochemical light-driven $CO_2$ conversion over the $PdSn_4$ single crystal. (b) EPR spectra of DMPO adducts over $Pd_{0.4}Pt_{0.6}Sn_4$ single crystal in the presence of $CO_2$. EPR spectra of DMPO adduct over $PdSn_4$ single crystal in (c) the absence and (d) presence of $CO_2$. Gibbs free energies of *H on the optimized surface of (e) $Pd_{0.4}Pt_{0.6}Sn_4$ and (f) $PdSn_4$ models. (g) In-situ DRIFTS spectra for light-driven $CO_2$ conversion over the $PdSn_4$ single crystal.

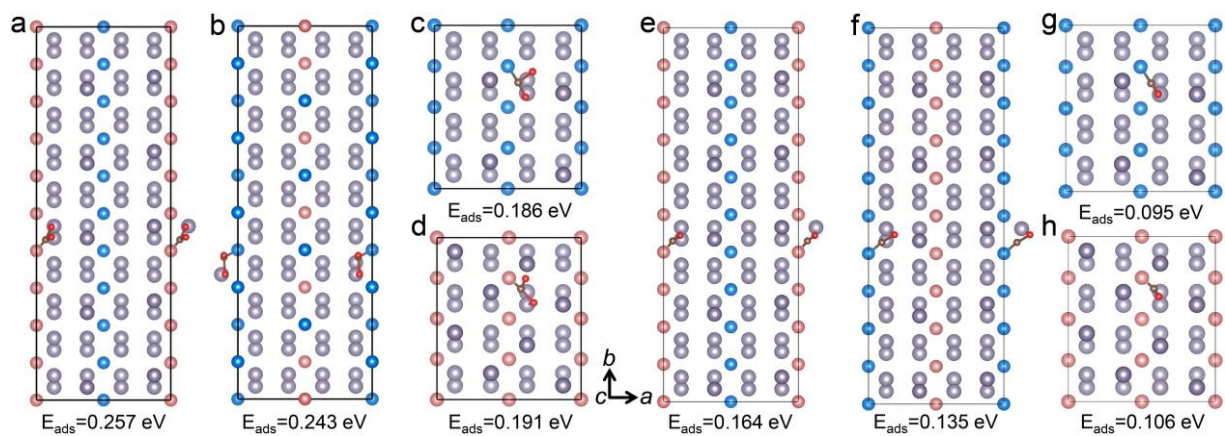

**Figure S13.** The crystal structures and atomic models of the (a) $Pd_{0.4}Pt_{0.6}Sn_4$ (010), (b) $Pd_{0.6}Pt_{0.4}Sn_4$ (010), (c) $PdSn_4$ (010), and (d) $PtSn_4$ (010) after adsorption of a $CO_2$ molecule. The crystal structures and atomic models of the (e) $Pd_{0.4}Pt_{0.6}Sn_4$ (010), (f) $Pd_{0.6}Pt_{0.4}Sn_4$ (010), (g) $PdSn_4$ (010), and (h) $PtSn_4$ (010) after adsorption of *CO.

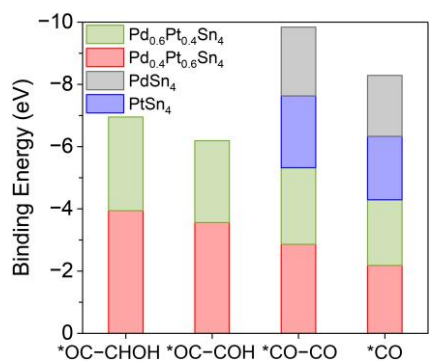

**Figure S14.** Calculated adsorption energies of *CO, *CO−CO, *OC−COH, and *OC−CHOH on the (010) surface of the $PdSn_4$, $PtSn_4$, $Pd_{0.4}Pt_{0.6}Sn_4$, and $Pd_{0.6}Pt_{0.4}Sn_4$ models.

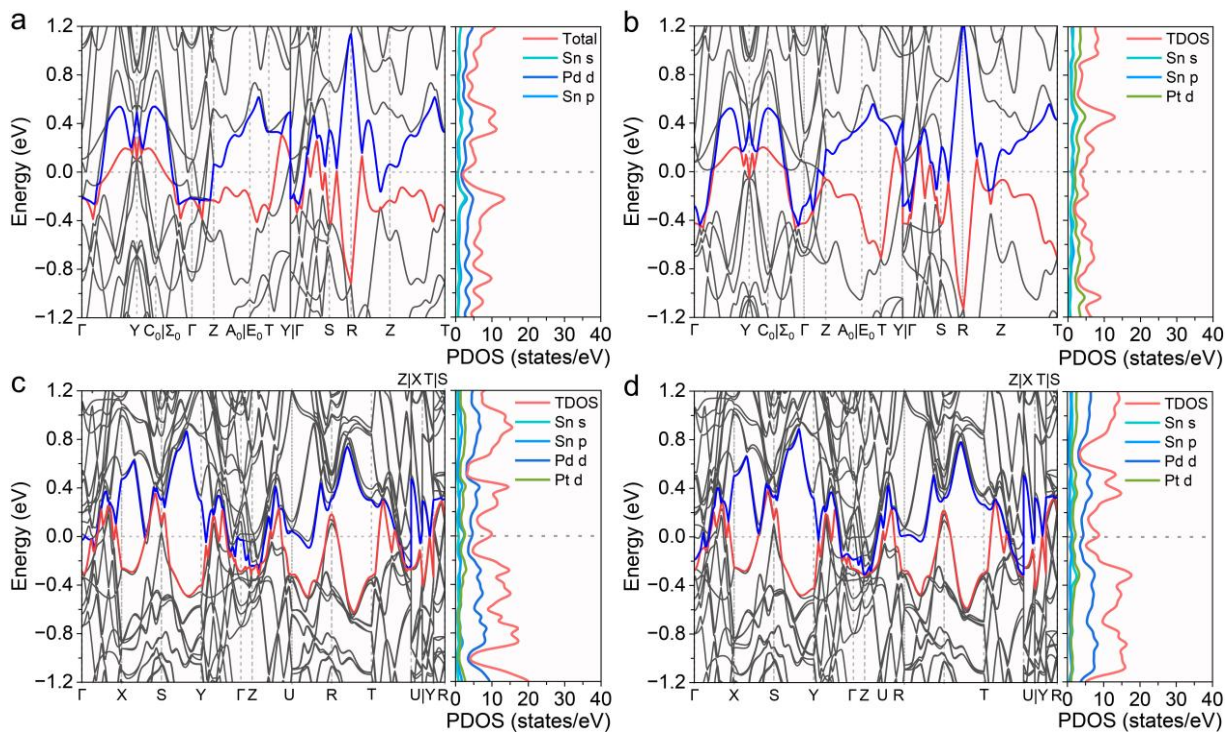

**Figure S15.** The bulk band structures and PDOS of the (a) PdSn$_4$, (b) PtSn$_4$, (c) Pd$_{0.4}$Pt$_{0.6}$Sn$_4$, and (d) Pd$_{0.6}$Pt$_{0.4}$Sn$_4$ models.

**Note 1: Electronic band structure and PDOS**

The bulk electronic band structure and PDOS of PdSn$_4$ and PtSn$_4$ models without SOC were calculated in detail; the few crossings and low DOS at the Fermi level ($E_F$) indicate the semi-metallic properties of the PdSn$_4$ and PtSn$_4$. Notably, it can be seen from the PDOS near the $E_F$ of PdSn$_4$ (Figure S15a, Supporting Information) and PtSn$_4$ (Figure S15b, Supporting Information) that the band is predominantly contributed by Pd *d* and Pt *d* orbitals, with a smaller contribution from Sn *s* and *p* orbitals. Due to the controllable electronic structure of transition metal d orbitals, it is possible to enhance the efficiency of photocatalytic reactions by regulating their electronic structure through doping. To optimize the surface electronic states of PdSn$_4$ for enhanced CRR performance, Pt atoms were doped into PdSn$_4$, replacing some Pd atoms to form Pd$_{0.4}$Pt$_{0.6}$Sn$_4$ and Pd$_{0.6}$Pt$_{0.4}$Sn$_4$. There were more band crossings near $E_F$, which modified the surface electronic structure of the catalyst to benefit the exposure of the active sites for CRR.[4]

In addition, the adsorption energy ($E_{ads}$) of $Pd_{0.6}Pt_{0.4}Sn_4$ decreased (Figure S14, Supporting Information), indicating a weaker interaction with $CO_2$, which may be unfavorable to the CRR.

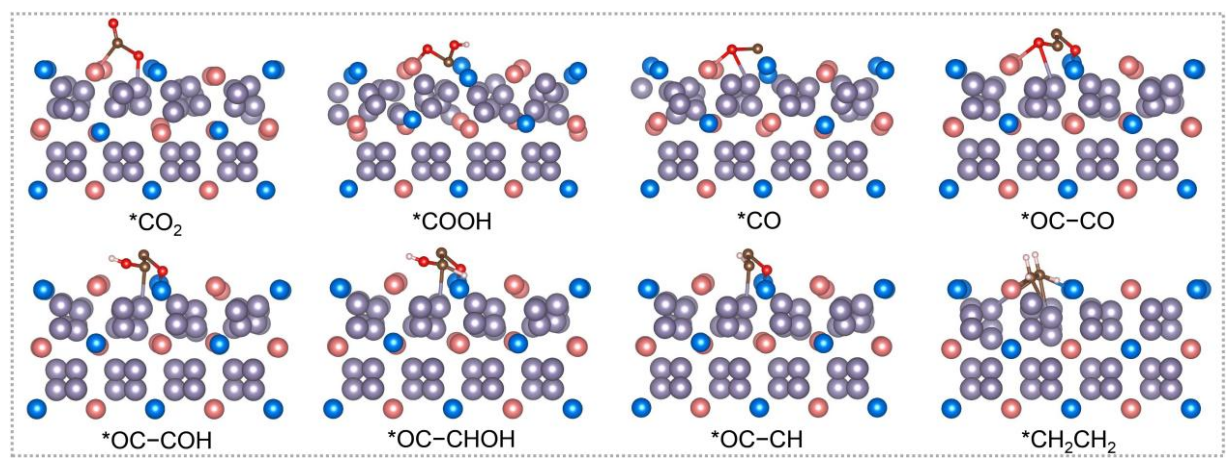

**Figure S16.** Schematic reaction CRR to $C_2H_4$ pathways projected on the optimized $Pd_{0.4}Pt_{0.6}Sn_4$ model surface (side view).

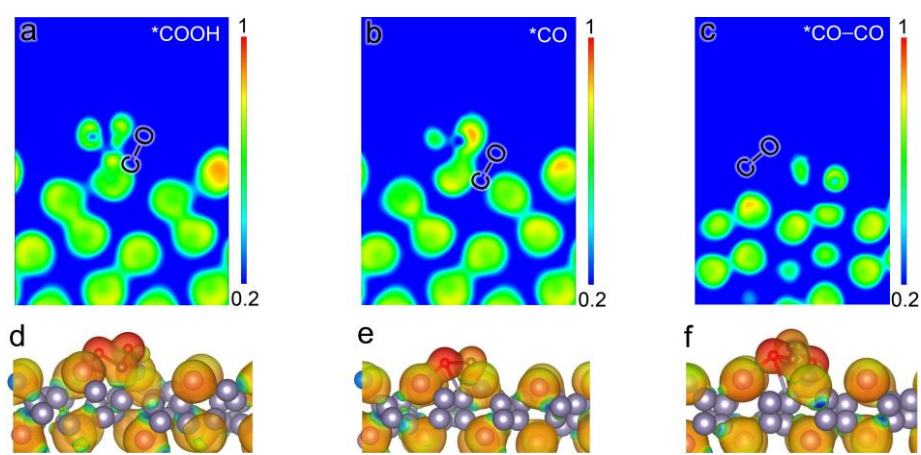

**Figure S17.** (a−c) The electron localization and (d−f) electrostatic potential for adsorption of *COOH, *CO, and *CO−CO on the surface of the optimized $Pd_{0.4}Pt_{0.6}Sn_4$ model.

**Table S1.** The average fluorescence lifetime of the $Pd_{0.4}Pt_{0.6}Sn_4$, $Pd_{0.6}Pt_{0.4}Sn_4$, $PdSn_4$, and $PtSn_4$ single crystals.

| Samples | $\tau_1$ (ns) | $B_1$ (%) | $\tau_2$ (ns) | $B_2$ (%) | Ave. $\tau$ (ns) |
|---|---|---|---|---|---|
| $Pd_{0.4}Pt_{0.6}Sn_4$ | 1.1148 | 1614.9711 | 6.6856 | 139.9642 | 3.0200 |
| $Pd_{0.6}Pt_{0.4}Sn_4$ | 1.2401 | 1627.0314 | 8.7004 | 92.1563 | 3.3617 |
| $PdSn_4$ | 1.6822 | 1401.7427 | 37.1069 | 70.8755 | 20.3602 |
| PtSn$_4$ | 1.3605 | 1473.5465 | 35.3885 | 56.5403 | 18.3581 |

**Table S2.** Carrier transport characteristics of the Pd$_{0.4}$Pt$_{0.6}$Sn$_4$, Pd$_{0.6}$Pt$_{0.4}$Sn$_4$, PdSn$_4$, and PtSn$_4$ single crystals at 300 K.[5]

| Materials | Height (m) | Hall coefficient (m$^3$ C$^{-1}$) | Carrier concentration (m$^{-3}$) | Mobility (m$^2$ V$^{-1}$ s$^{-1}$) | $L_D$ (μm) |
|---|---|---|---|---|---|
| Pd$_{0.4}$Pt$_{0.6}$Sn$_4$ | 5.9×10$^{-4}$ | 6.17×10$^{-11}$ | 1.01×10$^{29}$ | 8.36×10$^{-3}$ | 0.119−0.292 |
| Pd$_{0.6}$Pt$_{0.4}$Sn$_4$ | 3.2×10$^{-4}$ | 8.48×10$^{-11}$ | 7.36×10$^{28}$ | 5.32×10$^{-3}$ | 0.100−0.266 |
| PdSn$_4$ | 5.9×10$^{-4}$ | 3.56×10$^{-10}$ | 1.75×10$^{28}$ | 9.41×10$^{-4}$ | 0.049−0.231 |
| PtSn$_4$ | 1.2×10$^{-3}$ | 2.99×10$^{-10}$ | 2.09×10$^{28}$ | 1.02×10$^{-3}$ | 0.046−0.234 |

**Table S3.** Comparison of reduction paths of carbon products on the Pd$_{0.4}$Pt$_{0.6}$Sn$_4$ single crystal.

| C$_1$ products | C$_{2+}$ products |
|---|---|
| * + CO$_2$(g) + e$^-$ + H$^+$ → COOH*   (1) | CO* + e$^-$ + H$^+$ → *OCCO   (9) |
| COOH* + e$^-$ + H$^+$ → CO* + H$_2$O   (2) | *OCCO + e$^-$ + H$^+$ → *OCCOH   (10) |
| CO* + e$^-$ + H$^+$ → CHO* or CO* → CO(g)   (3) | *OCCOH + e$^-$ + H$^+$ → *OCCHOH   (11) |
| CHO* + e$^-$ + H$^+$ → CH$_2$O*   (4) | *OCCHOH + e$^-$ + H$^+$ → *OCCH   (12) |
| CH$_2$O* + e$^-$ + H$^+$ → CH$_3$O*   (5) | *OCCOH + e$^-$ + H$^+$ → *OCCH$_2$   (13) |
| CH$_3$O* + e$^-$ + H$^+$ → CH$_4$(g) + O*   (6) | *OCCH$_2$ + e$^-$ + H$^+$ → *OHCHCH$_2$   (14) |
| O* + e$^-$ + H$^+$ → OH*   (7) | *OHCHCH$_2$ + e$^-$ + H$^+$ → *CHCH$_2$   (15) |
| OH* + e$^-$ + H$^+$ → H$_2$O + *   (8) | *CHCH$_2$ + e$^-$ + H$^+$ → *CH$_2$CH$_2$   (16) |
|  | *CH$_2$CH$_2$ → CH$_2$CH$_2$(g) + *   (17) |

Where the "*" denotes the corresponding adsorption state on the surface of the single crystals.